\documentclass[preprint,prl,superscriptaddress,prbib,11pt]{revtex4}
\usepackage{graphicx}
\usepackage{bm}
\usepackage{hyperref}
\usepackage{todonotes}
\usepackage{nicefrac}
\usepackage{bbm}
\usepackage[normalem]{ulem}
\usepackage{booktabs}
\usepackage{subfigure}
\usepackage{rotating}
\usepackage{color}
\usepackage{float}
\usepackage{amsmath}
\usepackage[switch,columnwise]{lineno}
\begin{document}

\title{Room temperature observation of the anomalous in-plane Hall effect in epitaxial thin films of a Weyl ferromagnet}

\author{Soumya Sankar}
\affiliation{Department of Physics, The Hong Kong University of Science and Technology, Clear Water Bay, Kowloon, Hong Kong SAR}
\affiliation{These authors contributed equally}

\author{Xingkai Cheng}
\affiliation{Department of Physics, The Hong Kong University of Science and Technology, Clear Water Bay, Kowloon, Hong Kong SAR}
\affiliation{These authors contributed equally}

\author{Tahir Murtaza}
\affiliation{Department of Physics, The Hong Kong University of Science and Technology, Clear Water Bay, Kowloon, Hong Kong SAR}

\author{Caiyun Chen}
\affiliation{Department of Physics, The Hong Kong University of Science and Technology, Clear Water Bay, Kowloon, Hong Kong SAR}

\author{Yuqi Qin}
\affiliation{Department of Physics, The Hong Kong University of Science and Technology, Clear Water Bay, Kowloon, Hong Kong SAR}

\author{Xuezhao Wu}
\affiliation{Department of Electric and Computer Engineering, The Hong Kong University of Science and Technology, Clear Water Bay, Kowloon, Hong Kong SAR}

\author{Qiming Shao}
\affiliation{Department of Electric and Computer Engineering, The Hong Kong University of Science and Technology, Clear Water Bay, Kowloon, Hong Kong SAR}

\author{Rolf Lortz}
\affiliation{Department of Physics, The Hong Kong University of Science and Technology, Clear Water Bay, Kowloon, Hong Kong SAR}

\author{Junwei Liu}
\email[]{liuj@ust.hk}
\affiliation{Department of Physics, The Hong Kong University of Science and Technology, Clear Water Bay, Kowloon, Hong Kong SAR}

\author{Berthold J\"ack}
\email[]{bjaeck@ust.hk}
\affiliation{Department of Physics, The Hong Kong University of Science and Technology, Clear Water Bay, Kowloon, Hong Kong SAR}

\date{\today}

\begin{abstract}
Topologically nontrivial electronic states can give rise to novel anomalous Hall effects. The potential appearance of these effects at room temperature holds promise for their application in magnetic sensing, spintronics, and energy harvesting technology. The anomalous in-plane Hall effect (IPHE) is predicted to arise in topological magnetic materials when an external magnetic field is applied within the sample plane. Because of stringent symmetry requirements, the conclusive detection of the anomalous IPHE induced by topological electronic states remains challenging, and the study of anomalous Hall effects is often confined to cryogenic conditions. Combining molecular beam epitaxy of the kagome metal Fe$_3$Sn with measurements of the electric Hall effect and theoretical calculations, we propose and experimentally demonstrate that the interplay of the kagome lattice motif with spin-orbit coupling and canted ferromagnetism with large exchange interactions gives rise to the anomalous IPHE at room temperature that is induced by topological Weyl points in the electronic band structure. Synthesizing a topological heterostructure including layers of Fe$_3$Sn and ferromagnetic CoFeB, we further show the enhancement of the anomalous IPHE through the magnetic stray field of the CoFeB layer. Our work establishes a design paradigm for topological magnets and heterostructures to discover and control novel anomalous Hall effects toward their use in technological applications.

\end{abstract}

\maketitle

Topologically nontrivial electronic states can give rise to novel electric transport phenomena, such as linear and nonlinear anomalous Hall effects~\cite{nagaosa2010anomalous, ma2019observation, gao2023quantum, wang2023quantum, sankar2024experimental}. These phenomena provide deep insight into the fundamental properties of topological quantum materials and have garnered increasing attention for their applications in magnetic sensing, spintronics, and energy harvesting technology~\cite{nakatsuji2015large, ni2016ultrahigh, onishi2024high}. Their experimental detection requires that the total Berry curvature of the occupied electronic states does not vanish. This can be practically realized in magnetic materials and heterostructures in which the magnetic structure breaks the time-reversal symmetry $\mathcal{T}$, resulting in finite anomalous Hall conductivity~\cite{nagaosa2010anomalous, sodemann2015quantum, zhang2023higher}. As long as the energy scale of this symmetry breaking mechanism, such as the magnetic exchange interaction $J$, is sufficiently large, Berry curvature effects on electric transport properties can be detected even at and above room temperature~\cite{ye2018massive, sankar2024experimental}, favoring their potential integration in technological applications. To date, the observation of novel anomalous Hall effects remains often limited to cryogenic temperatures~\cite{liang2018anomalous, ma2019observation, lai2021third, gao2023quantum, wang2023quantum}. This motivates the search for new classes of topological magnetic materials and heterostructures, which can be manufactured using scalable synthesis methods, to discover and control Hall effects at room temperature.

The anomalous in-plane Hall effect (IPHE) describes the appearance of an electric Hall voltage when an external magnetic field is applied within the sample plane~\cite{liu2013plane, sun2022possible}. It contrasts with the conventional Hall effects that can be observed when a magnetic field applied along the out-of-plane direction. The IPHE is predicted to arise from the finite Berry curvature of topologically nontrivial electronic states of magnetic materials. It has more stringent symmetry requirements than the conventional anomalous Hall effect (requires broken $\mathcal{T}$) and requires the additional breaking of crystalline symmetries such as out-of-plane rotation symmetries~\cite{liu2013plane, sun2022possible}. Because these types of symmetries are present in most crystalline materials, the observation of the anomalous IPHE remains a challenging task. To date, IPHEs are mostly field-induced, where the breaking of the relevant crystalline symmetries is realized by an external magnetic field~\cite{liang2018anomalous, wang2024orbital, nakamura2024observation}; moreover, an anomalous IPHE independent of temperature, resulting from topological electronic states, has not yet been reported~\cite{zhou2022heterodimensional}.

Combining electric Hall measurements of epitaxially grown thin films of the Weyl ferromagnet Fe$_3$Sn and systematic theoretical calculations, we propose and experimentally demonstrate that the interplay of the kagome lattice motif with spin-orbit coupling and canted in-plane ferromagnetism with large exchange interactions $J\approx40\,$meV~\cite{sales2019electronic} gives rise to the anomalous IPHE at room temperature. Our observation of a temperature-independent anomalous in-plane Hall conductivity over a wide temperature window between 100\,K and 300\,K indicates that this effect is induced by topological Weyl points near the Fermi energy~\cite{belbase2023large}. Synthesizing a topological heterostructure that comprises layers of Fe$_3$Sn and ferromagnetic CoFeB, we also demonstrate the tuning of the IPHE amplitude by $\approx38\,\%$. Utilizing thin film growth of topological kagome magnets and heterostructures with large magnetic exchange energies, our study thus establishes a design paradigm for the discovery of novel Hall effects at room temperature toward their technological application.

\subsection{Prediction of the anomalous in-plane Hall effect in the Weyl ferromagnet Fe$_3$Sn}

The kagome metal Fe$_3$Sn ($a = b = 5.487\,$Å and $c = 4.3\,$Å, space group $P6_3/mmc$) belongs to the class of binary kagome metals $X_{\rm p} Y_{\rm q}$, which attracts a lot of interest due to its topological and flat band electronic states ($X=\text{Fe, Co, Ni}$, $Y=\text{Sn, In, Ge}$, with chemical stoichiometry $p,\,q$)~\cite{ye2018massive, yin2018giant, liu2018giant, kang2020dirac, kang2020topological, liu2020orbital, sankar2024experimental, chen2023visualizing, chen2024cascade}. Common among this material class are strong nearest-neighbor magnetic exchange interactions between transition metal atoms arranged on the kagome lattice, as seen in Fig.~\ref{fig:fig1}(a) that can result in a long-range magnetic order much above room temperature~\cite{ye2018massive, sales2019electronic, prodan2023large}. 

In Fe$_3$Sn, the Fe ions form a kagome lattice in the crystallographic $ab$ plane and a Sn ion occupy the honeycomb center, as seen in Fig.~\ref{fig:fig1}(a). The primitive unit cell of Fe$_3$Sn is composed of kagome bilayers stacked along the crystallographic $[0001]$-direction. For simplicity, we introduce the Cartesian coordinates $x$, $y$, and $z$ corresponding to the crystalline directions $[2\bar{1}\bar{1}0]$, $[01\bar{1}0]$, and [0001], respectively. An easy-plane magneto-crystalline anisotropy imparts in-plane ferromagnetism with the magnetzation $\bf M$ pointing along the $x$ direction, where large ferromagnetic exchange interactions $J\approx40\,$meV~\cite{sales2019electronic} results in a Curie temperature $T_{\rm C}\approx705$\,K exceeding room temperature~\cite{belbase2023large, prodan2023large, kurosawa2024large, fu2024magnetic, prodan2024anisotropic}.

Ferromagnetism combined with spin-orbit coupling profoundly influences the electronic structure and topology of kagome metals~\cite{ye2018massive, kang2020dirac, sankar2024experimental}. In Fe$_3$Sn, this combination imparts a set of Weyl points when the magnetization vector points along the $x$ direction~\cite{belbase2023large}. An additional magnetization component along the $z$ direction, such as resulting from a canting of the magnetic moments of Fe, further lifts characteristic degeneracies of the electronic band structure of the kagome lattice, such as occurring at the crystallographic $\Gamma$ and $K$ points of the hexagonal Brioullin zone. This results in a set of Weyl points near the Fermi energy~\cite{belbase2023large}. Using symmetry analyses and {\em ab-initio} model calculations, we show that these Weyl points give rise to an anomalous in-plane Hall conductivity $\sigma_{\rm IPHE}$ in spin-canted Fe$_3$Sn.

Given that the magnetization vector is oriented along the $x$-direction, the magnetic space group (MSG) of Fe$_3$Sn is $Cmc^{\prime}m^{\prime}$, which consists of three mirror (glide-mirror) symmetries $\mathcal{M}_{\rm x}$, $\mathcal{M}_{\rm y}t\mathcal{T}$, and $\mathcal{M}_{\rm z}t\mathcal{T}$, where $t$ is a fractional translation. Determined by all the symmetries in the MSG, the total Berry curvature (as seen in Fig.~\ref{fig:fig1}, (b) and (d)) and thus the anomalous Hall conductivity $\sigma_{\rm xy}$ within the sample $xy$-plane vanishes; only an out-of-plane component $\sigma_{\rm yz}$ is permitted (see Sec.~I of the Suppl.~Materials for more details). On the other hand, an out-of-plane magnetic field or out-of-plane canting of the magnetization along the $z$-direction breaks $\mathcal{M}_{\rm x}$ and $\mathcal{M}_{\rm z}t\mathcal{T}$ and permits $\sigma_{\rm xy}\neq0$, since the total Berry curvature $\Omega_{\rm z}$ is finite in this case (as seen in Fig.~\ref{fig:fig1}, (c) and (e)). 

We also analyzed the influence of canting on the the electronic band structure of Fe$_3$Sn by carrying out {\em ab initio} calculations (see the Methods section for calculation details). While the total Berry curvature vanishes in the absence of canting (as seen in Fig.~\ref{fig:fig1}(d)), the presence of canting ${\bf M}=(M_{\rm x},\,0,\,M_{\rm z})$ results in a finite Berry curvature (as seen in Fig.~\ref{fig:fig1}(e)). Therefore, our symmetry analysis and model calculations propose that Fe$_3$Sn should allow the observation of the anomalous IPHE when finite spin canting in the $z$ direction is present. An in-plane magnetic field $\bf{B}_{\parallel}= \rm{B}\left(\cos{\alpha},\,\sin{\alpha},\,0\right)$ can further modulate the anomalous in-plane Hall conductivity $\sigma_{\rm IPHE}$ through the exchange coupling between $\bf{B}_{\parallel}$ and the spin of the itinerant electrons. Here, $\alpha$ corresponds to the azimuthal angle between $\bf M$ and $\bf{B}_{\parallel}$. We have analyzed this effect using a Wannier tight-binding model of the electronic band structure of Fe$_3$Sn (see the Methods section for details). The resulting $\sigma_{\rm IPHE}(\alpha)$ is shown in Fig.~\ref{fig:fig1}(f). Our calculations establish that $\bf{B}_{\parallel}$ modulates $\sigma_{\rm IPHE}$ with $2\pi$ periodicity in the presence of canting. As we show below, this distinct dependence of $\sigma_{\rm IPHE}$ on the relative angle between $\bf{B}_{\parallel}$ and $\bf M$ allows us to separate the IPHE from other contributions to the in-plane Hall response.


\subsection{Experimental detection of spin-canted in-plane ferromagnetism in Fe$_3$Sn thin films}

We developed the molecular beam epitaxy on Fe$_3$Sn thin films (typical thicknesses 30-60\,nm) on the (0001) surface of sapphire substrates using a Pt(111) buffer layer of 5\,nm thickness to facilitate layer-by-layer growth of continuous films (see the Methods section and Sec.~II of the Suppl.~Materials for results of the chemical and structural characterization). Transmission electron microscopy measurements reveal a defined Pt(111)-Fe$_3$Sn interface and confirm the characteristic bilayer stacking of the Fe$_3$Sn kagome layers along the $z$ direction, as seen in Fig.~\ref{fig:fig2}(a). Measurements of the temperature ($T$) dependent longitudinal resistivity shown in Fig.~\ref{fig:fig2}(b) demonstrate the metallic character of the Fe$_3$Sn thin films. The residual $\rho_{xx}\approx72\times10^{-6}\,\Omega\,\text{cm}$ at $T=2\,$K agrees well with previously reported values for single crystals~\cite{belbase2023large, prodan2023large, prodan2024anisotropic}, indicating a high crystalline quality and a low defect density of our samples. Note that the Pt(111) buffer layer is metallic and could shunt the electric conduction through the Fe$_3$Sn layer in current-biased measurements. Our analysis of this shunt effect, discussed in Sec.~III of the Suppl.~Materials, shows that electric conduction through the Pt(111) buffer layer has only a modest quantitative effect on the measured transverse voltages at all temperatures and will be neglected for what follows.

The results of our magnetization measurements at $T=300\,$K as a function of an externally applied magnetic field $B$ are shown in Fig.~\ref{fig:fig2}(c). In agreement with previous results on sputtered thin films and bulk crystals~\cite{belbase2023large, prodan2023large, kurosawa2024large}, we detect an anisotropic response to an externally applied magnetic field ${\bf B}$; the saturation field is smaller when ${\bf B}$ is applied within the $xy$-plane than when ${\bf B}$ is applied along the $z$-axis. This result is consistent with the presence of in-plane ferromagnetism along the $x$-axis. The detected magnetic moment of $\approx2.3\,\mu_{\rm B}/\text{Fe atom}$ matches well with previous reports~\cite{belbase2023large, prodan2023large, kurosawa2024large} and is dominated by the spin moment of the Fe atoms~\cite{prodan2023large}. A closer examination of the magnetization curve at small out-of-plane fields further reveals the presence of a narrow hysteresis loop with remnant out-of-plane magnetization $M_{\rm z}\approx0.11\,\mu_{\rm B}/\text{Fe atom}$. This observation indicates the presence of a finite out-of-plane canting of the magnetic easy-plane of Fe in our thin films. 

Fe$_3$Sn films were shaped into circular Hall bar devices, as seen in Fig.~\ref{fig:fig2}(d) (see Methods section for details of the device fabrication). This device geometry allows us to characterize the complete $4\pi$ angle dependence of the in-plane Hall response with respect to the crystallographic $y$ axis, the electric field $\bf E$, and the in-plane magnetic field $\bf B_{\parallel}$, as defined in Fig.~\ref{fig:fig2}(d). Here, $\bf E$ lies in the sample plane and is always parallel to the direction of the applied bias current $I$ (see Methods section for measurement details). Specifically, we will present results from electric transport measurements as a function of the azimuthal angle $\theta$ defined as the angle between $\bf E$ and $\bf B_{\parallel}$, and the azimuthal angle $\phi$ defined as the angle between $\bf E$ and the $y$ axis of the sample. The magnitude of $\phi$ increases in a counterclockwise direction, as defined in Fig.~\ref{fig:fig2}(d), and we fix $\phi=0$ along the $y$-direction. For example, $\phi=0$ and $\phi=\pi/2$ correspond to $I$ applied from contacts 1 to 7 and 4 to 10, respectively. Moreover, we will also present results for measurements in an out-of-plane geometry with the magnetic field aligned along the $z$ direction. For clarity in what follows, we introduce the polar angle $\Omega$ defined as the angle between $\bf B$ and the $z$ direction (see Fig.~\ref{fig:fig2}(d)). Therefore, $\Omega=0$ and $\Omega=\pi/2$ for ${\bf B}\parallel z$ and ${\bf B}\perp z$, respectively.


We first characterize the conventional Hall response of a Fe$_3$Sn thin film (thickness $d_{\rm 1}=60\,$nm) at room temperature when the magnetic field ${\bf B}$ is applied along the z-direction ($\Omega=0$). The Hall resistivity $\rho_{\rm H}(B,\,\phi=0)$ is shown in Fig.~\ref{fig:fig2}(e). It corresponds to the $\bf B$-antisymmetric contribution to the transverse resistivity $\rho_{\rm xy}$ and can be obtained by means of the symmetrization operation $\rho_{\rm H}(B,\,\phi)=(\rho_{\rm xy}(B,\,\phi,\,\Omega=0)+\rho_{\rm xy}(B,\,\phi+\pi/2,\,\Omega=0))/2$. The symmetric contribution to $\rho_{\rm xy}$ is discussed in Sec.~IV of the Suppl.~Materials. As can be seen, $\rho_{\rm H}(B)\propto B$ in small fields and saturates at $\rho_{\rm H}(B)\approx5\times10^{-6}\,\Omega$ cm at $B>1\,$ T, consistent with the results of magnetization measurements ({\em c.f.}~Fig.~\ref{fig:fig1}(c)). These observations agree well with previous results~\cite{belbase2023large,prodan2023large,kurosawa2024large}. This saturation effect was reported to be associated with the magnetic field-induced reorientation of Fe magnetic moments from the $x$-axis to the $z$-axis~\cite{prodan2023large}. Examining $\rho_{\rm xy}(B,\,\phi=0,\,\Omega=0)$ over a smaller field range for upward and downward sweeps of $B$ reveals the presence of a hysteresis loop, as seen in Fig.~\ref{fig:fig2}(f). The observation of hysteresis in both the orbital Hall and magnetization measurements experimentally support the presence of an out-of-plane canting of the Fe magnetic moments at zero applied magnetic field that should permit the observation of the anomalous IPHE by breaking the $\mathcal{M}_{\rm z}t\mathcal{T}$ symmetry.

\subsection{Observation of the in-plane Hall effect at room temperature}

In the next step, we characterize the Hall response of Fe$_3$Sn at $T=300\,$K when the magnetic field is applied within the sample plane ($\Omega=\pi/2$). To this end, we record the transverse resistivity $\rho_{\rm xy}(\phi,\,\theta,\,\Omega=\pi/2, B_{\parallel})$ as a function of $\phi$ and $\theta$ at $T=300\,K$ and $B_{\parallel}=\pm20\,$mT. Using symmetrization operations $\rho_{\rm IPHE}(\phi,\,\theta)=(\rho_{\rm xy}(\phi,\,\theta,\,\Omega=\pi/2), B_{\parallel})-\rho_{\rm xy}(\phi,\,\theta,\,\Omega=\pi/2), -B_{\parallel})/2$ and $\rho_{\rm sym}(\phi,\,\theta)=(\rho_{\rm xy}(\phi,\,\theta,\,\Omega=\pi/2), B_{\parallel})+\rho_{\rm xy}(\phi,\,\theta,\,\Omega=\pi/2), -B_{\parallel})/2$, we obtain the antisymmetric and symmetric contributions to $\rho_{\rm xy}(\phi,\,\theta,\,\Omega=\pi/2, B_{\parallel})$, respectively.

We first consider the antisymmetric part $\rho_{\rm IPHE}(\phi,\,\theta)$, which is a true Hall effect, because it is antis-symmetric in $B_{\parallel}$. Figure~\ref{fig:fig3}(a) displays $\rho_{\rm IPHE}(\phi,\,\theta)$ for device D1 at $\phi=0$, $\phi=\pi/2$, $\phi=\pi$, and $\phi=3\pi/2$, that is for different angles $\phi$ between the electric field and the $y$ axis of the sample. The complete $4\pi$ angle dependence of $\rho_{\rm IPHE}(\phi,\,\theta)$ in the $(\phi,\,\theta)$ plane is shown in Fig.~\ref{fig:fig3}(b). We detect a characteristic $2\pi$-periodicity of $\rho_{\rm IPHE}(\phi,\,\theta)$ with respect to $\theta$ at all $\phi$-values. Moreover, $\rho_{\rm IPHE}(\theta)$ recorded at $\phi=0$ ($\phi=\pi/2$) and $\phi=\pi$ ($\phi=3\pi/2$) exhibit a $\pi$-shift with respect to $\theta$. As shown in Fig.~\ref{fig:fig3}(a), we can fit the $\theta$-dependence of $\rho_{\rm IPHE}(\phi,\,\theta)$ with $\rho_{\rm IPHE}(\phi,\,\theta)=\rho_{\rm AHE}+\rho_{\rm IPHE}^0 \sin(\theta+\phi+\delta)$, where $\delta$ accounts for small angular offsets; the fit parameters are listed in Sec.~V of the Suppl.~Materials. Note that these fits reveal the absence of a sizable vertical offset $\rho_{\rm AHE}$, as shown in the inset of Fig.~\ref{fig:fig3}(a) ($\rho_{\rm AHE}/max(\rho_{\rm IPHE})<1\,\%$). This indicates that the $\theta$ independent contribution to $\rho_{\rm AHE}$ owing to the breaking of the $M_{\rm x}$ symmetry by the canting of the magnetic moments of Fe along the $z$ direction is negligible in our measurements. 

The $2\pi$ periodicity and phase shift of $\rho_{\rm IPHE}(\phi,\,\theta)$ with respect to $\theta$ shown in Fig.~\ref{fig:fig3}, (a) and (b) can be understood in the context of our results from {\em ab initio} calculations. Here, the IPHE is modulated with $2\pi$ periodicity by the exchange coupling between $\bf B_{\parallel}$ and the spins of the itinerant electrons, depending on the relative angle between $\bf B_{\parallel}$ and $\bf M$, as seen in Fig.~\ref{fig:fig1}(f)). Let us first assume that $\bf B_{\parallel}$ is always parallel to $\bf E$ in our experiment ($\theta=0$). Since $\bf M$ is parallel to the direction $x$ in our sample, $\bf B_{\parallel}$ and $\bf M$ thus cover different angles from $\phi=0$ to $\phi=2\pi$ when $\bf E$ is applied along different contacts of the circular Hall bar device. Given the $2\pi$ periodicity of the IPHE modulation by $\bf B_{\parallel}$ seen in our calculations, it follows that $\rho_{\rm IPHE}(\phi=0,\,\theta=0)=-\rho_{\rm IPHE}(\phi=\pi,\,\theta=0)$, as observed in our measurements. Second, consider the scenario in which $\bf E$ is applied only along one contact pair, for example $\phi=0$, and the angle $\theta$ between $\bf E$ and $\bf B_{\parallel}$ can be varied. Here, a complete $2\pi$ rotation of $\bf B_{\parallel}$ always covers a complete $2\pi$ rotation of the relative angle between $\bf B_{\parallel}$ and $\bf M$. Therefore, a $2\pi$ rotation of $\theta$ results in a $2\pi$ periodic modulation of $\rho_{\rm IPHE}(\theta)$ at any given value of $\phi$, as observed in our measurements.

Next, we consider the full $4\pi$ angle dependence of the symmetric contribution $\rho_{\rm sym}(\phi,\,\theta)$ to the in-plane Hall response in the $(\phi,\,\theta)$ plane, as shown in Fig.~\ref{fig:fig3}(c). The circular Hall bar geometry allows us to decompose $\rho_{\rm sym}(\phi,\,\theta)$ into two separate contributions, $\rho_{\rm STR}(\phi)$ and $\rho_{\rm PHE}(\theta)$ through their separate dependencies on $\phi$ and $\theta$, as clearly indicated in Fig.~\ref{fig:fig3}(c). Here, $\rho_{\rm STR}(\phi)$, shown in Fig.~\ref{fig:fig3}(d), only depends on $\phi$ and describes the symmetric transverse resistivity (STR). The STR is $\pi$ periodic in $\phi$ and can be understood to arise from the angular misalignment $\phi$ of $\bf E$ with the principal axes of Fe$_3$Sn in the presence of in-plane magnetism (see Sec.~IV of Suppl.~Materials for a detailed analysis). Hence, STR is a projection of the anisotropic magnetoresistance effect, which commonly appears in longitudinal resistivity, into transverse resistivity~\cite{li2022anisotropic}. The second contribution $\rho_{\rm PHE}(\theta)$, shown in Fig.~\ref{fig:fig3}(e), describes the so-called planar Hall effect (PHE), which exhibits a $\pi$-periodicity with respect to $\theta$~\cite{li2022anisotropic}. Note that the PHE is not a true Hall effect because it is magnetic-field symmetric. At the same time, $\rho_{\rm PHE}(\theta)$ is an antisymmetric tensor and $\phi$-invariant. Thus, it can be distinguished from $\rho_{\rm STR}(\phi)$. The detected $\pi$-periodicity of $\rho_{\rm PHE}(\theta)$ can be described by a $\pi$-periodic sine function and is a manifestation of the remaining $C_{\rm 2y}$-symmetry of ferromagnetic Fe$_3$Sn. Finally, the regular periodicity of $\rho_{\rm IPHE}(\phi,\,\theta)$ and $\rho_{\rm sym}(\phi,\,\theta)$ within the $(\phi,\,\theta)$-plane suggests that the effect of magnetic domains, if present, on the in-plane Hall response is negligible.

\subsection{Origin of the in-plane Hall effect}

Our theoretical analysis suggests that the experimentally detected canting of the Fe magnetic moments permits an anomalous IPHE by breaking the relevant magnetic symmetry ($\mathcal{M}_{\rm z}t\mathcal{T}$). Therefore, we expect the IPHE amplitude to be independent of the amplitude of $\bf B_{\parallel}$, which only modulates $\rho_{\rm IPHE}(\phi,\,\theta)$ as a function of $\theta$. To examine this relation, we recorded the response of the IPHE to different amplitudes of $\bf B_{\parallel}$, as shown in Fig.~\ref{fig:fig4}(a). A $2\pi$-periodic IPHE can be detected even in fields as small as 5\,mT. Moreover, the IPHE amplitude remains nearly field-independent at $B_{\parallel}<100\,$ mT, as seen in in the inset of Fig.~\ref{fig:fig4}(a). This observation supports our understanding that the magnetic field is not responsible for breaking $\mathcal{M}_{\rm z}t\mathcal{T}$. Hence, $\mathcal{M}_{\rm z}t\mathcal{T}$ is already broken by the canting of the magnetic moments, experimentally supporting the observation of the anomalous IPHE. 

Only at $B_{\parallel}\geq100\,$mT, the IPHE amplitude increases monotonically with increasing magnetic field amplitude and the $\theta$ dependence exhibits a continuous phase shift of $\pi$. This effect can probably be attributed to an additional external contribution to $\rho_{\rm IPHE}(\phi,\,\theta)$ by an orbital Hall effect (OHE). This OHE appear in the presence of a minute angular misalignment of the magnetic field direction with the $xy$ plane of the sample causing a finite out-of-plane component of the magnetic field~\cite{liang2018anomalous, wang2024orbital}. Under the $2\pi$ rotation of the sample holder in a magnetic field with fixed direction, this misalignment causes a sine-like modulation of the OHE with $2\pi$ periodicity with respect to $\theta$. Since the OHE is magnetic field antisymmetric, it thus can contribute a field-dependent part to $\rho_{\rm IPHE}(\theta)$ at $B_{\parallel}\geq100\,$ mT.

The experimental characteristics of in-plane Hall measurements of spin-canted Fe$_3$Sn are consistent with our theoretical proposal of an anomalous IPHE induced by topological Weyl points in the kagome band structure, as shown in Fig.~\ref{fig:fig1}(e). Berry curvature contributions to the Hall conductivity are generally independent of the charge carrier scattering time. They can thus be distinguished from other contributions, such as skew scattering and side jumps~\cite{smit1955spontaneous, smit1958spontaneous, berger1970side, nagaosa2010anomalous}, which depend on the scattering time. Accordingly, Berry curvature contributions to the Hall effect can be identified by analyzing the temperature dependence of the anomalous Hall conductivity.

To examine this relation, we characterize the IPHE as a function of the temperature. The resulting anomalous in-plane Hall conductivity $\sigma_{\rm IPHE}(T,\,\phi=\pi,\,\theta=\pi/2,\,)$, as well as the longitudinal conductivity $\sigma_{\rm xx}(T)$ are shown in Fig.~\ref{fig:fig4}(b). As can be seen, the amplitude of $\sigma_{\rm IPHE}(T,\,\phi=\pi,\,\theta=\pi/2)$ is temperature independent at $T\geq100\,$ K but increases at $T<100\,$K. These characteristics contrast with those of $\sigma_{\rm xx}(T)$, which depends on $T$ throughout the examined temperature range. The experimental observation of a temperature-independent contribution $|\sigma_{\rm IPHE}^0|\approx0.8\,\text{S\,cm}^{-1}$ to the IPHE suggests that the Hall response at $T\geq100\,$K is dominated by an intrinsic Berry curvature contribution~\cite{nagaosa2010anomalous}, consistent with our expectations from symmetry analyses and model calculations. Moreover, the experimentally detected $|\sigma_{\rm IPHE}^0|\approx0.8\,\text{S\,cm}^{-1}$ is also in relatively good quantitative agreement with the calculated in-plane Hall conductivity $|\sigma_{\rm IPHE}^{\rm calc}|\approx1.5\,\text{S\,cm}^{-1}$, as shown in Fig.~\ref{fig:fig1}(f). At lower temperatures, the scaling $\sigma_{\rm IPHE}(T)\propto\sigma_{\rm xx}$, as seen in Fig.~\ref{fig:fig4}(c), indicates that the Hall conductivity is dominated by skew-scattering contributions~\cite{smit1955spontaneous, smit1958spontaneous, nagaosa2010anomalous}.

Breaking inversion symmetry at the Fe$_3$Sn/Pt(111) interface could induce a canting of the Fe magnetic moments near the interface via the Dzyaloshinskii-Moriya (DM) interaction~\cite{hellman2017interface}. To test whether the experimentally observed canting in our samples originates from this effect, we examined in the in-plane Hall response of thin films with smaller thickness ($d_{\rm 2}=30\,$nm). If the IPHE arises from interfacial symmetry-breaking effects, the corresponding antisymmetric part $V_{\rm IPHE}(\theta)$ of the Hall voltage $V_{\rm xy}(\theta)$ is independent of the film thickness. The $V_{\rm IPHE}(\theta,\,\phi=0)$ of devices D1 ($d_{\rm 1}=60\,\text{nm}$) and D2 ($d_{\rm 2}=30\,\text{nm}$) are shown in Fig.~\ref{fig:fig4}(d). Applying a sine function fit as before, we find $V_{\rm IPHE}(60\,\text{nm})/V_{\rm IPHE}(30\,\text{nm})\approx2.5$, i.e., $V_{\rm IPHE}$ scales approximately with the thickness of the film. Since $\rho_{\rm xy}\propto V_{\rm xy}d$ ($d$ is the thickness of the film), it follows that $\rho_{\rm IPHE}(\phi,\,\theta)$ and $\rho_{\rm sym}(\phi,\,\theta)$ of devices D1 and D2 are in qualitative and approximate quantitative agreement ($\rho_{\rm IPHE}(\phi,\,\theta)$ and $\rho_{\rm sym}(\phi,\,\theta)$ of device D2 are shown in Sec.~VI of the Suppl.~Materials). 

Our results suggest that the IPHE is independent of the film thickness, and thus arises from spin canting within the bulk of the film. Since our films are comparably thick ($d\leq60\,$nm), it is difficult to reconcile the fact that epitaxial strain at the Pt(111)-Fe$_3$Sn interface gives rise to the detected canting effect. Although it remains an interesting question to which extent this canting is generally present in Fe$_3$Sn~\cite{prodan2023large, prodan2024anisotropic}, it has been reported that the magnetization of the iso-structural ferromagnet Fe$_3$Ge undergoes a reorientation from an out-of-plane to an in-plane magnetization via a second-order transition below 380\,K~\cite{drijver1976magnetic, lou2024orbital}. It is conceivable that such a transition is present in Fe$_3$Sn, leading to finite out-of-plane canting at finite temperatures. Thus, it would be desirable to conduct more temperature-dependent neutron scattering measurements on bulk crystals to obtain more detailed information on the magnetic structure of Fe$_3$Sn.

\subsection{Tuning the in-plane Hall effect in a topological heterostructure}

The amplitude of a Hall effect induced by Berry curvature generally depends on the magnitude of the underlying mechanism of symmetry breaking~\cite{nagaosa2010anomalous}. Tuning the spin-canting in Fe$_3$Sn through external control parameters would thus allow the control of the anomalous IPHE. We will show that the magnetic stray field of ferromagnetic CoFeB layers augments the IPHE amplitude via the coupling to the magnetic moments of Fe. To this end, we realize a topological heterostructure structure comprising of TiO$_2$\,(3\,nm)/CoFeB\,(20\,nm)/Al$_2$O$_3$\,(20\,nm)/Fe$_3$Sn\,(30\,nm)/Pt(111)\,(5\,nm) layers deposited on sapphire(0001), as schematically shown in Fig.~\ref{fig:fig5}(a) (see Methods section for fabrication details). The electrically isolating Al$_2$O$_3$ spacer prevents electric shunting of Fe$_3$Sn by the metallic CoFeB. The successful fabrication of the topological heterostructure is confirmed by scanning transmission electron microscopy and energy-dispersive X-ray spectroscopy measurements shown in Fig.~\ref{fig:fig5}, (b)-(d) (see Methods section for measurement details). In addition, we performed a detailed characterization of the crystallographic and magnetic properties of the CoFeB films, as shown in Sec.~VII of the Suppl.~Materials. These measurements indicate that CoFeB exhibits an amorphous crystal structure with a tilted easy-plane magnetic anisotropy. Hence, the CoFeB films of device D2{\em hs} are expected to have a magnetic stray field with a component in the $z$ direction.

To maintain direct comparability of the measured IPHE amplitudes, we deposited the (TiO$_2$/CoFeB/Al$_2$O$_3$)-layer structure directly on the circular Hall bar device D2 using a two-step lithography process. We then recorded the electric in-plane Hall response of device D2{\em hs} ({\em hs} - heterostructure). The resulting $\rho_{\rm IPHE}(\phi=0,\,\theta)$ of the devices D2{\em hs} and D2 are shown in Fig.~\ref{fig:fig5}(e). We find that the amplitude of the IPHE is increased by approximately 38$\,\%$ after the deposition of the CoFeB film. The $2\pi$-periodicity and $\phi$-dependence of the IPHE remain unchanged, as can be seen from the full $(\phi,\,\theta)$-dependence of $\rho_{\rm IPHE}$ and $\rho_{\rm sym}$ shown in Sec.~VII.D of the Suppl.~Materials. On the other hand, the symmetric part $\rho_{\rm sym}(\phi,\,\theta)$ of the Hall resistivity of device D2{\em hs} and D2 are in qualitative and quantitative agreement. At a basic level and disregarding effects from the microscopic magnetic structure of CoFeB, these observations suggest that the CoFeB layer leaves the in-plane magnetic structure of Fe$_3$ Sn invariant, but enhances the out-of-plane canting effect through the coupling of the magnetic moments of Fe to its magnetic stray field. We anticipate that further optimization of the heterostructure layers and the use of other kagome materials~\cite{ye2018massive, wang2018large} could enhance the IPHE amplitude even further beyond what was detected in this proof-of-concept experiment. 

\subsection{Conclusion}
In conclusion, our study reports the experimental observation of the anomalous IPHE in epitaxially grown thin films of the Weyl ferromagnet Fe$_3$Sn at room temperature. To this end, the combination of the kagome lattice motif with spin orbit coupling and ferromagnetism imparts a topological electronic structure~\cite{belbase2023large} with broken time-reversal symmetry. Meanwhile, strong ferromagnetic exchange interactions $J\approx40\,$meV between the transition metal atoms arranged on the kagome lattice~\cite{sales2019electronic} enable the observation at room temperature of this phenomenon. The temperature-independent anomalous in-plane Hall conductivity detected over a wide temperature range (100-300\,K) provides strong evidence for the topological origin of the anomalous IPHE, consistent with the presence of topological Weyl points near Fermi energy. Establishing the thin-film heteroepitaxy of topological kagome magnets and heterostructures, our work establishes a design paradigm to examine new electric Hall effects, such as anomalous and nonlinear Hall effects~\cite{sodemann2015quantum, ma2019observation, gao2023quantum, wang2023quantum, zhang2023higher, sankar2024experimental} at room temperature toward their use in technological applications, such as magnetic sensors, spintronics devices, and energy harvesting.

\bibliography{bibliography}

\begin{thebibliography}{47}
\expandafter\ifx\csname natexlab\endcsname\relax\def\natexlab#1{#1}\fi
\expandafter\ifx\csname bibnamefont\endcsname\relax
  \def\bibnamefont#1{#1}\fi
\expandafter\ifx\csname bibfnamefont\endcsname\relax
  \def\bibfnamefont#1{#1}\fi
\expandafter\ifx\csname citenamefont\endcsname\relax
  \def\citenamefont#1{#1}\fi
\expandafter\ifx\csname url\endcsname\relax
  \def\url#1{\texttt{#1}}\fi
\expandafter\ifx\csname urlprefix\endcsname\relax\def\urlprefix{URL }\fi
\providecommand{\bibinfo}[2]{#2}
\providecommand{\eprint}[2][]{\url{#2}}

\bibitem[{\citenamefont{Nagaosa et~al.}(2010{\natexlab{a}})\citenamefont{Nagaosa, Sinova, Onoda, MacDonald, and Ong}}]{nagaosa2010anomalous}
\bibinfo{author}{\bibfnamefont{N.}~\bibnamefont{Nagaosa}}, \bibinfo{author}{\bibfnamefont{J.}~\bibnamefont{Sinova}}, \bibinfo{author}{\bibfnamefont{S.}~\bibnamefont{Onoda}}, \bibinfo{author}{\bibfnamefont{A.~H.} \bibnamefont{MacDonald}}, \bibnamefont{and} \bibinfo{author}{\bibfnamefont{N.~P.} \bibnamefont{Ong}}, \bibinfo{journal}{Reviews of modern physics} \textbf{\bibinfo{volume}{82}}, \bibinfo{pages}{1539} (\bibinfo{year}{2010}{\natexlab{a}}).

\bibitem[{\citenamefont{Ma et~al.}(2019)\citenamefont{Ma, Xu, Shen, MacNeill, Fatemi, Chang, Mier~Valdivia, Wu, Du, Hsu et~al.}}]{ma2019observation}
\bibinfo{author}{\bibfnamefont{Q.}~\bibnamefont{Ma}}, \bibinfo{author}{\bibfnamefont{S.-Y.} \bibnamefont{Xu}}, \bibinfo{author}{\bibfnamefont{H.}~\bibnamefont{Shen}}, \bibinfo{author}{\bibfnamefont{D.}~\bibnamefont{MacNeill}}, \bibinfo{author}{\bibfnamefont{V.}~\bibnamefont{Fatemi}}, \bibinfo{author}{\bibfnamefont{T.-R.} \bibnamefont{Chang}}, \bibinfo{author}{\bibfnamefont{A.~M.} \bibnamefont{Mier~Valdivia}}, \bibinfo{author}{\bibfnamefont{S.}~\bibnamefont{Wu}}, \bibinfo{author}{\bibfnamefont{Z.}~\bibnamefont{Du}}, \bibinfo{author}{\bibfnamefont{C.-H.} \bibnamefont{Hsu}}, \bibnamefont{et~al.}, \bibinfo{journal}{Nature} \textbf{\bibinfo{volume}{565}}, \bibinfo{pages}{337} (\bibinfo{year}{2019}).

\bibitem[{\citenamefont{Gao et~al.}(2023)\citenamefont{Gao, Liu, Qiu, Ghosh, V.~Trevisan, Onishi, Hu, Qian, Tien, Chen et~al.}}]{gao2023quantum}
\bibinfo{author}{\bibfnamefont{A.}~\bibnamefont{Gao}}, \bibinfo{author}{\bibfnamefont{Y.-F.} \bibnamefont{Liu}}, \bibinfo{author}{\bibfnamefont{J.-X.} \bibnamefont{Qiu}}, \bibinfo{author}{\bibfnamefont{B.}~\bibnamefont{Ghosh}}, \bibinfo{author}{\bibfnamefont{T.}~\bibnamefont{V.~Trevisan}}, \bibinfo{author}{\bibfnamefont{Y.}~\bibnamefont{Onishi}}, \bibinfo{author}{\bibfnamefont{C.}~\bibnamefont{Hu}}, \bibinfo{author}{\bibfnamefont{T.}~\bibnamefont{Qian}}, \bibinfo{author}{\bibfnamefont{H.-J.} \bibnamefont{Tien}}, \bibinfo{author}{\bibfnamefont{S.-W.} \bibnamefont{Chen}}, \bibnamefont{et~al.}, \bibinfo{journal}{Science} \textbf{\bibinfo{volume}{381}}, \bibinfo{pages}{181} (\bibinfo{year}{2023}).

\bibitem[{\citenamefont{Wang et~al.}(2023)\citenamefont{Wang, Kaplan, Zhang, Holder, Cao, Wang, Zhou, Zhou, Jiang, Zhang et~al.}}]{wang2023quantum}
\bibinfo{author}{\bibfnamefont{N.}~\bibnamefont{Wang}}, \bibinfo{author}{\bibfnamefont{D.}~\bibnamefont{Kaplan}}, \bibinfo{author}{\bibfnamefont{Z.}~\bibnamefont{Zhang}}, \bibinfo{author}{\bibfnamefont{T.}~\bibnamefont{Holder}}, \bibinfo{author}{\bibfnamefont{N.}~\bibnamefont{Cao}}, \bibinfo{author}{\bibfnamefont{A.}~\bibnamefont{Wang}}, \bibinfo{author}{\bibfnamefont{X.}~\bibnamefont{Zhou}}, \bibinfo{author}{\bibfnamefont{F.}~\bibnamefont{Zhou}}, \bibinfo{author}{\bibfnamefont{Z.}~\bibnamefont{Jiang}}, \bibinfo{author}{\bibfnamefont{C.}~\bibnamefont{Zhang}}, \bibnamefont{et~al.}, \bibinfo{journal}{Nature} \textbf{\bibinfo{volume}{621}}, \bibinfo{pages}{487} (\bibinfo{year}{2023}).

\bibitem[{\citenamefont{Sankar et~al.}(2024)\citenamefont{Sankar, Liu, Zhang, Li, Chen, Gao, Zheng, Lin, Qian, Yu et~al.}}]{sankar2024experimental}
\bibinfo{author}{\bibfnamefont{S.}~\bibnamefont{Sankar}}, \bibinfo{author}{\bibfnamefont{R.}~\bibnamefont{Liu}}, \bibinfo{author}{\bibfnamefont{C.-P.} \bibnamefont{Zhang}}, \bibinfo{author}{\bibfnamefont{Q.-F.} \bibnamefont{Li}}, \bibinfo{author}{\bibfnamefont{C.}~\bibnamefont{Chen}}, \bibinfo{author}{\bibfnamefont{X.-J.} \bibnamefont{Gao}}, \bibinfo{author}{\bibfnamefont{J.}~\bibnamefont{Zheng}}, \bibinfo{author}{\bibfnamefont{Y.-H.} \bibnamefont{Lin}}, \bibinfo{author}{\bibfnamefont{K.}~\bibnamefont{Qian}}, \bibinfo{author}{\bibfnamefont{R.-P.} \bibnamefont{Yu}}, \bibnamefont{et~al.}, \bibinfo{journal}{Physical Review X} \textbf{\bibinfo{volume}{14}}, \bibinfo{pages}{021046} (\bibinfo{year}{2024}).

\bibitem[{\citenamefont{Nakatsuji et~al.}(2015)\citenamefont{Nakatsuji, Kiyohara, and Higo}}]{nakatsuji2015large}
\bibinfo{author}{\bibfnamefont{S.}~\bibnamefont{Nakatsuji}}, \bibinfo{author}{\bibfnamefont{N.}~\bibnamefont{Kiyohara}}, \bibnamefont{and} \bibinfo{author}{\bibfnamefont{T.}~\bibnamefont{Higo}}, \bibinfo{journal}{Nature} \textbf{\bibinfo{volume}{527}}, \bibinfo{pages}{212} (\bibinfo{year}{2015}).

\bibitem[{\citenamefont{Ni et~al.}(2016)\citenamefont{Ni, Zhang, Nlebedim, and Jiles}}]{ni2016ultrahigh}
\bibinfo{author}{\bibfnamefont{Y.}~\bibnamefont{Ni}}, \bibinfo{author}{\bibfnamefont{Z.}~\bibnamefont{Zhang}}, \bibinfo{author}{\bibfnamefont{I.}~\bibnamefont{Nlebedim}}, \bibnamefont{and} \bibinfo{author}{\bibfnamefont{D.~C.} \bibnamefont{Jiles}}, \bibinfo{journal}{IEEE Transactions on Magnetics} \textbf{\bibinfo{volume}{52}}, \bibinfo{pages}{1} (\bibinfo{year}{2016}).

\bibitem[{\citenamefont{Onishi and Fu}(2024)}]{onishi2024high}
\bibinfo{author}{\bibfnamefont{Y.}~\bibnamefont{Onishi}} \bibnamefont{and} \bibinfo{author}{\bibfnamefont{L.}~\bibnamefont{Fu}}, \bibinfo{journal}{Physical Review B} \textbf{\bibinfo{volume}{110}}, \bibinfo{pages}{075122} (\bibinfo{year}{2024}).

\bibitem[{\citenamefont{Sodemann and Fu}(2015)}]{sodemann2015quantum}
\bibinfo{author}{\bibfnamefont{I.}~\bibnamefont{Sodemann}} \bibnamefont{and} \bibinfo{author}{\bibfnamefont{L.}~\bibnamefont{Fu}}, \bibinfo{journal}{Physical review letters} \textbf{\bibinfo{volume}{115}}, \bibinfo{pages}{216806} (\bibinfo{year}{2015}).

\bibitem[{\citenamefont{Zhang et~al.}(2023)\citenamefont{Zhang, Gao, Xie, Po, and Law}}]{zhang2023higher}
\bibinfo{author}{\bibfnamefont{C.-P.} \bibnamefont{Zhang}}, \bibinfo{author}{\bibfnamefont{X.-J.} \bibnamefont{Gao}}, \bibinfo{author}{\bibfnamefont{Y.-M.} \bibnamefont{Xie}}, \bibinfo{author}{\bibfnamefont{H.~C.} \bibnamefont{Po}}, \bibnamefont{and} \bibinfo{author}{\bibfnamefont{K.~T.} \bibnamefont{Law}}, \bibinfo{journal}{Physical Review B} \textbf{\bibinfo{volume}{107}}, \bibinfo{pages}{115142} (\bibinfo{year}{2023}).

\bibitem[{\citenamefont{Ye et~al.}(2018)\citenamefont{Ye, Kang, Liu, Von~Cube, Wicker, Suzuki, Jozwiak, Bostwick, Rotenberg, Bell et~al.}}]{ye2018massive}
\bibinfo{author}{\bibfnamefont{L.}~\bibnamefont{Ye}}, \bibinfo{author}{\bibfnamefont{M.}~\bibnamefont{Kang}}, \bibinfo{author}{\bibfnamefont{J.}~\bibnamefont{Liu}}, \bibinfo{author}{\bibfnamefont{F.}~\bibnamefont{Von~Cube}}, \bibinfo{author}{\bibfnamefont{C.~R.} \bibnamefont{Wicker}}, \bibinfo{author}{\bibfnamefont{T.}~\bibnamefont{Suzuki}}, \bibinfo{author}{\bibfnamefont{C.}~\bibnamefont{Jozwiak}}, \bibinfo{author}{\bibfnamefont{A.}~\bibnamefont{Bostwick}}, \bibinfo{author}{\bibfnamefont{E.}~\bibnamefont{Rotenberg}}, \bibinfo{author}{\bibfnamefont{D.~C.} \bibnamefont{Bell}}, \bibnamefont{et~al.}, \bibinfo{journal}{Nature} \textbf{\bibinfo{volume}{555}}, \bibinfo{pages}{638} (\bibinfo{year}{2018}).

\bibitem[{\citenamefont{Liang et~al.}(2018)\citenamefont{Liang, Lin, Gibson, Kushwaha, Liu, Wang, Xiong, Sobota, Hashimoto, Kirchmann et~al.}}]{liang2018anomalous}
\bibinfo{author}{\bibfnamefont{T.}~\bibnamefont{Liang}}, \bibinfo{author}{\bibfnamefont{J.}~\bibnamefont{Lin}}, \bibinfo{author}{\bibfnamefont{Q.}~\bibnamefont{Gibson}}, \bibinfo{author}{\bibfnamefont{S.}~\bibnamefont{Kushwaha}}, \bibinfo{author}{\bibfnamefont{M.}~\bibnamefont{Liu}}, \bibinfo{author}{\bibfnamefont{W.}~\bibnamefont{Wang}}, \bibinfo{author}{\bibfnamefont{H.}~\bibnamefont{Xiong}}, \bibinfo{author}{\bibfnamefont{J.~A.} \bibnamefont{Sobota}}, \bibinfo{author}{\bibfnamefont{M.}~\bibnamefont{Hashimoto}}, \bibinfo{author}{\bibfnamefont{P.~S.} \bibnamefont{Kirchmann}}, \bibnamefont{et~al.}, \bibinfo{journal}{Nature Physics} \textbf{\bibinfo{volume}{14}}, \bibinfo{pages}{451} (\bibinfo{year}{2018}).

\bibitem[{\citenamefont{Lai et~al.}(2021)\citenamefont{Lai, Liu, Zhang, Zhao, Feng, Wang, Tang, Liu, Novoselov, Yang et~al.}}]{lai2021third}
\bibinfo{author}{\bibfnamefont{S.}~\bibnamefont{Lai}}, \bibinfo{author}{\bibfnamefont{H.}~\bibnamefont{Liu}}, \bibinfo{author}{\bibfnamefont{Z.}~\bibnamefont{Zhang}}, \bibinfo{author}{\bibfnamefont{J.}~\bibnamefont{Zhao}}, \bibinfo{author}{\bibfnamefont{X.}~\bibnamefont{Feng}}, \bibinfo{author}{\bibfnamefont{N.}~\bibnamefont{Wang}}, \bibinfo{author}{\bibfnamefont{C.}~\bibnamefont{Tang}}, \bibinfo{author}{\bibfnamefont{Y.}~\bibnamefont{Liu}}, \bibinfo{author}{\bibfnamefont{K.}~\bibnamefont{Novoselov}}, \bibinfo{author}{\bibfnamefont{S.~A.} \bibnamefont{Yang}}, \bibnamefont{et~al.}, \bibinfo{journal}{Nature Nanotechnology} \textbf{\bibinfo{volume}{16}}, \bibinfo{pages}{869} (\bibinfo{year}{2021}).

\bibitem[{\citenamefont{Liu et~al.}(2013)\citenamefont{Liu, Hsu, and Liu}}]{liu2013plane}
\bibinfo{author}{\bibfnamefont{X.}~\bibnamefont{Liu}}, \bibinfo{author}{\bibfnamefont{H.-C.} \bibnamefont{Hsu}}, \bibnamefont{and} \bibinfo{author}{\bibfnamefont{C.-X.} \bibnamefont{Liu}}, \bibinfo{journal}{Physical review letters} \textbf{\bibinfo{volume}{111}}, \bibinfo{pages}{086802} (\bibinfo{year}{2013}).

\bibitem[{\citenamefont{Sun et~al.}(2022)\citenamefont{Sun, Weng, and Dai}}]{sun2022possible}
\bibinfo{author}{\bibfnamefont{S.}~\bibnamefont{Sun}}, \bibinfo{author}{\bibfnamefont{H.}~\bibnamefont{Weng}}, \bibnamefont{and} \bibinfo{author}{\bibfnamefont{X.}~\bibnamefont{Dai}}, \bibinfo{journal}{Physical Review B} \textbf{\bibinfo{volume}{106}}, \bibinfo{pages}{L241105} (\bibinfo{year}{2022}).

\bibitem[{\citenamefont{Wang et~al.}(2024)\citenamefont{Wang, Zhu, Chen, Wang, Liu, Huang, Jiang, Zhao, Shi, Tian et~al.}}]{wang2024orbital}
\bibinfo{author}{\bibfnamefont{L.}~\bibnamefont{Wang}}, \bibinfo{author}{\bibfnamefont{J.}~\bibnamefont{Zhu}}, \bibinfo{author}{\bibfnamefont{H.}~\bibnamefont{Chen}}, \bibinfo{author}{\bibfnamefont{H.}~\bibnamefont{Wang}}, \bibinfo{author}{\bibfnamefont{J.}~\bibnamefont{Liu}}, \bibinfo{author}{\bibfnamefont{Y.-X.} \bibnamefont{Huang}}, \bibinfo{author}{\bibfnamefont{B.}~\bibnamefont{Jiang}}, \bibinfo{author}{\bibfnamefont{J.}~\bibnamefont{Zhao}}, \bibinfo{author}{\bibfnamefont{H.}~\bibnamefont{Shi}}, \bibinfo{author}{\bibfnamefont{G.}~\bibnamefont{Tian}}, \bibnamefont{et~al.}, \bibinfo{journal}{Physical Review Letters} \textbf{\bibinfo{volume}{132}}, \bibinfo{pages}{106601} (\bibinfo{year}{2024}).

\bibitem[{\citenamefont{Nakamura et~al.}(2024)\citenamefont{Nakamura, Nishihaya, Ishizuka, Kriener, Watanabe, and Uchida}}]{nakamura2024observation}
\bibinfo{author}{\bibfnamefont{A.}~\bibnamefont{Nakamura}}, \bibinfo{author}{\bibfnamefont{S.}~\bibnamefont{Nishihaya}}, \bibinfo{author}{\bibfnamefont{H.}~\bibnamefont{Ishizuka}}, \bibinfo{author}{\bibfnamefont{M.}~\bibnamefont{Kriener}}, \bibinfo{author}{\bibfnamefont{Y.}~\bibnamefont{Watanabe}}, \bibnamefont{and} \bibinfo{author}{\bibfnamefont{M.}~\bibnamefont{Uchida}}, \bibinfo{journal}{Physical Review Letters} \textbf{\bibinfo{volume}{133}}, \bibinfo{pages}{236602} (\bibinfo{year}{2024}).

\bibitem[{\citenamefont{Zhou et~al.}(2022)\citenamefont{Zhou, Zhang, Lin, Cao, Zhou, Jiang, Du, Tang, Shi, Jiang et~al.}}]{zhou2022heterodimensional}
\bibinfo{author}{\bibfnamefont{J.}~\bibnamefont{Zhou}}, \bibinfo{author}{\bibfnamefont{W.}~\bibnamefont{Zhang}}, \bibinfo{author}{\bibfnamefont{Y.-C.} \bibnamefont{Lin}}, \bibinfo{author}{\bibfnamefont{J.}~\bibnamefont{Cao}}, \bibinfo{author}{\bibfnamefont{Y.}~\bibnamefont{Zhou}}, \bibinfo{author}{\bibfnamefont{W.}~\bibnamefont{Jiang}}, \bibinfo{author}{\bibfnamefont{H.}~\bibnamefont{Du}}, \bibinfo{author}{\bibfnamefont{B.}~\bibnamefont{Tang}}, \bibinfo{author}{\bibfnamefont{J.}~\bibnamefont{Shi}}, \bibinfo{author}{\bibfnamefont{B.}~\bibnamefont{Jiang}}, \bibnamefont{et~al.}, \bibinfo{journal}{Nature} \textbf{\bibinfo{volume}{609}}, \bibinfo{pages}{46} (\bibinfo{year}{2022}).

\bibitem[{\citenamefont{Sales et~al.}(2019)\citenamefont{Sales, Yan, Meier, Christianson, Okamoto, and McGuire}}]{sales2019electronic}
\bibinfo{author}{\bibfnamefont{B.~C.} \bibnamefont{Sales}}, \bibinfo{author}{\bibfnamefont{J.}~\bibnamefont{Yan}}, \bibinfo{author}{\bibfnamefont{W.~R.} \bibnamefont{Meier}}, \bibinfo{author}{\bibfnamefont{A.~D.} \bibnamefont{Christianson}}, \bibinfo{author}{\bibfnamefont{S.}~\bibnamefont{Okamoto}}, \bibnamefont{and} \bibinfo{author}{\bibfnamefont{M.~A.} \bibnamefont{McGuire}}, \bibinfo{journal}{Physical Review Materials} \textbf{\bibinfo{volume}{3}}, \bibinfo{pages}{114203} (\bibinfo{year}{2019}).

\bibitem[{\citenamefont{Belbase et~al.}(2023)\citenamefont{Belbase, Ye, Karki, Facio, You, Checkelsky, Van Den~Brink, and Ghimire}}]{belbase2023large}
\bibinfo{author}{\bibfnamefont{B.~P.} \bibnamefont{Belbase}}, \bibinfo{author}{\bibfnamefont{L.}~\bibnamefont{Ye}}, \bibinfo{author}{\bibfnamefont{B.}~\bibnamefont{Karki}}, \bibinfo{author}{\bibfnamefont{J.~I.} \bibnamefont{Facio}}, \bibinfo{author}{\bibfnamefont{J.-S.} \bibnamefont{You}}, \bibinfo{author}{\bibfnamefont{J.~G.} \bibnamefont{Checkelsky}}, \bibinfo{author}{\bibfnamefont{J.}~\bibnamefont{Van Den~Brink}}, \bibnamefont{and} \bibinfo{author}{\bibfnamefont{M.~P.} \bibnamefont{Ghimire}}, \bibinfo{journal}{Physical Review B} \textbf{\bibinfo{volume}{108}}, \bibinfo{pages}{075164} (\bibinfo{year}{2023}).

\bibitem[{\citenamefont{Yin et~al.}(2018)\citenamefont{Yin, Zhang, Li, Jiang, Chang, Zhang, Lian, Xiang, Belopolski, Zheng et~al.}}]{yin2018giant}
\bibinfo{author}{\bibfnamefont{J.-X.} \bibnamefont{Yin}}, \bibinfo{author}{\bibfnamefont{S.~S.} \bibnamefont{Zhang}}, \bibinfo{author}{\bibfnamefont{H.}~\bibnamefont{Li}}, \bibinfo{author}{\bibfnamefont{K.}~\bibnamefont{Jiang}}, \bibinfo{author}{\bibfnamefont{G.}~\bibnamefont{Chang}}, \bibinfo{author}{\bibfnamefont{B.}~\bibnamefont{Zhang}}, \bibinfo{author}{\bibfnamefont{B.}~\bibnamefont{Lian}}, \bibinfo{author}{\bibfnamefont{C.}~\bibnamefont{Xiang}}, \bibinfo{author}{\bibfnamefont{I.}~\bibnamefont{Belopolski}}, \bibinfo{author}{\bibfnamefont{H.}~\bibnamefont{Zheng}}, \bibnamefont{et~al.}, \bibinfo{journal}{Nature} \textbf{\bibinfo{volume}{562}}, \bibinfo{pages}{91} (\bibinfo{year}{2018}).

\bibitem[{\citenamefont{Liu et~al.}(2018)\citenamefont{Liu, Sun, Kumar, Muechler, Sun, Jiao, Yang, Liu, Liang, Xu et~al.}}]{liu2018giant}
\bibinfo{author}{\bibfnamefont{E.}~\bibnamefont{Liu}}, \bibinfo{author}{\bibfnamefont{Y.}~\bibnamefont{Sun}}, \bibinfo{author}{\bibfnamefont{N.}~\bibnamefont{Kumar}}, \bibinfo{author}{\bibfnamefont{L.}~\bibnamefont{Muechler}}, \bibinfo{author}{\bibfnamefont{A.}~\bibnamefont{Sun}}, \bibinfo{author}{\bibfnamefont{L.}~\bibnamefont{Jiao}}, \bibinfo{author}{\bibfnamefont{S.-Y.} \bibnamefont{Yang}}, \bibinfo{author}{\bibfnamefont{D.}~\bibnamefont{Liu}}, \bibinfo{author}{\bibfnamefont{A.}~\bibnamefont{Liang}}, \bibinfo{author}{\bibfnamefont{Q.}~\bibnamefont{Xu}}, \bibnamefont{et~al.}, \bibinfo{journal}{Nature physics} \textbf{\bibinfo{volume}{14}}, \bibinfo{pages}{1125} (\bibinfo{year}{2018}).

\bibitem[{\citenamefont{Kang et~al.}(2020{\natexlab{a}})\citenamefont{Kang, Ye, Fang, You, Levitan, Han, Facio, Jozwiak, Bostwick, Rotenberg et~al.}}]{kang2020dirac}
\bibinfo{author}{\bibfnamefont{M.}~\bibnamefont{Kang}}, \bibinfo{author}{\bibfnamefont{L.}~\bibnamefont{Ye}}, \bibinfo{author}{\bibfnamefont{S.}~\bibnamefont{Fang}}, \bibinfo{author}{\bibfnamefont{J.-S.} \bibnamefont{You}}, \bibinfo{author}{\bibfnamefont{A.}~\bibnamefont{Levitan}}, \bibinfo{author}{\bibfnamefont{M.}~\bibnamefont{Han}}, \bibinfo{author}{\bibfnamefont{J.~I.} \bibnamefont{Facio}}, \bibinfo{author}{\bibfnamefont{C.}~\bibnamefont{Jozwiak}}, \bibinfo{author}{\bibfnamefont{A.}~\bibnamefont{Bostwick}}, \bibinfo{author}{\bibfnamefont{E.}~\bibnamefont{Rotenberg}}, \bibnamefont{et~al.}, \bibinfo{journal}{Nature materials} \textbf{\bibinfo{volume}{19}}, \bibinfo{pages}{163} (\bibinfo{year}{2020}{\natexlab{a}}).

\bibitem[{\citenamefont{Kang et~al.}(2020{\natexlab{b}})\citenamefont{Kang, Fang, Ye, Po, Denlinger, Jozwiak, Bostwick, Rotenberg, Kaxiras, Checkelsky et~al.}}]{kang2020topological}
\bibinfo{author}{\bibfnamefont{M.}~\bibnamefont{Kang}}, \bibinfo{author}{\bibfnamefont{S.}~\bibnamefont{Fang}}, \bibinfo{author}{\bibfnamefont{L.}~\bibnamefont{Ye}}, \bibinfo{author}{\bibfnamefont{H.~C.} \bibnamefont{Po}}, \bibinfo{author}{\bibfnamefont{J.}~\bibnamefont{Denlinger}}, \bibinfo{author}{\bibfnamefont{C.}~\bibnamefont{Jozwiak}}, \bibinfo{author}{\bibfnamefont{A.}~\bibnamefont{Bostwick}}, \bibinfo{author}{\bibfnamefont{E.}~\bibnamefont{Rotenberg}}, \bibinfo{author}{\bibfnamefont{E.}~\bibnamefont{Kaxiras}}, \bibinfo{author}{\bibfnamefont{J.~G.} \bibnamefont{Checkelsky}}, \bibnamefont{et~al.}, \bibinfo{journal}{Nature communications} \textbf{\bibinfo{volume}{11}}, \bibinfo{pages}{4004} (\bibinfo{year}{2020}{\natexlab{b}}).

\bibitem[{\citenamefont{Liu et~al.}(2020)\citenamefont{Liu, Li, Wang, Wang, Wen, Jiang, Lu, Yan, Huang, Shen et~al.}}]{liu2020orbital}
\bibinfo{author}{\bibfnamefont{Z.}~\bibnamefont{Liu}}, \bibinfo{author}{\bibfnamefont{M.}~\bibnamefont{Li}}, \bibinfo{author}{\bibfnamefont{Q.}~\bibnamefont{Wang}}, \bibinfo{author}{\bibfnamefont{G.}~\bibnamefont{Wang}}, \bibinfo{author}{\bibfnamefont{C.}~\bibnamefont{Wen}}, \bibinfo{author}{\bibfnamefont{K.}~\bibnamefont{Jiang}}, \bibinfo{author}{\bibfnamefont{X.}~\bibnamefont{Lu}}, \bibinfo{author}{\bibfnamefont{S.}~\bibnamefont{Yan}}, \bibinfo{author}{\bibfnamefont{Y.}~\bibnamefont{Huang}}, \bibinfo{author}{\bibfnamefont{D.}~\bibnamefont{Shen}}, \bibnamefont{et~al.}, \bibinfo{journal}{Nature communications} \textbf{\bibinfo{volume}{11}}, \bibinfo{pages}{4002} (\bibinfo{year}{2020}).

\bibitem[{\citenamefont{Chen et~al.}(2023)\citenamefont{Chen, Zheng, Yu, Sankar, Law, Po, and J{\"a}ck}}]{chen2023visualizing}
\bibinfo{author}{\bibfnamefont{C.}~\bibnamefont{Chen}}, \bibinfo{author}{\bibfnamefont{J.}~\bibnamefont{Zheng}}, \bibinfo{author}{\bibfnamefont{R.}~\bibnamefont{Yu}}, \bibinfo{author}{\bibfnamefont{S.}~\bibnamefont{Sankar}}, \bibinfo{author}{\bibfnamefont{K.~T.} \bibnamefont{Law}}, \bibinfo{author}{\bibfnamefont{H.~C.} \bibnamefont{Po}}, \bibnamefont{and} \bibinfo{author}{\bibfnamefont{B.}~\bibnamefont{J{\"a}ck}}, \bibinfo{journal}{Physical Review Research} \textbf{\bibinfo{volume}{5}}, \bibinfo{pages}{043269} (\bibinfo{year}{2023}).

\bibitem[{\citenamefont{Chen et~al.}(2024)\citenamefont{Chen, Zheng, He, Ying, Sankar, Li, Wei, Dai, Po, and J{\"a}ck}}]{chen2024cascade}
\bibinfo{author}{\bibfnamefont{C.}~\bibnamefont{Chen}}, \bibinfo{author}{\bibfnamefont{J.}~\bibnamefont{Zheng}}, \bibinfo{author}{\bibfnamefont{Y.}~\bibnamefont{He}}, \bibinfo{author}{\bibfnamefont{X.}~\bibnamefont{Ying}}, \bibinfo{author}{\bibfnamefont{S.}~\bibnamefont{Sankar}}, \bibinfo{author}{\bibfnamefont{L.}~\bibnamefont{Li}}, \bibinfo{author}{\bibfnamefont{Y.}~\bibnamefont{Wei}}, \bibinfo{author}{\bibfnamefont{X.}~\bibnamefont{Dai}}, \bibinfo{author}{\bibfnamefont{H.~C.} \bibnamefont{Po}}, \bibnamefont{and} \bibinfo{author}{\bibfnamefont{B.}~\bibnamefont{J{\"a}ck}}, \bibinfo{journal}{arXiv preprint arXiv:2409.06933}  (\bibinfo{year}{2024}).

\bibitem[{\citenamefont{Prodan et~al.}(2023)\citenamefont{Prodan, Evans, Griffin, {\"O}stlin, Altthaler, Lysne, Filippova, Shova, Chioncel, Tsurkan et~al.}}]{prodan2023large}
\bibinfo{author}{\bibfnamefont{L.}~\bibnamefont{Prodan}}, \bibinfo{author}{\bibfnamefont{D.~M.} \bibnamefont{Evans}}, \bibinfo{author}{\bibfnamefont{S.~M.} \bibnamefont{Griffin}}, \bibinfo{author}{\bibfnamefont{A.}~\bibnamefont{{\"O}stlin}}, \bibinfo{author}{\bibfnamefont{M.}~\bibnamefont{Altthaler}}, \bibinfo{author}{\bibfnamefont{E.}~\bibnamefont{Lysne}}, \bibinfo{author}{\bibfnamefont{I.~G.} \bibnamefont{Filippova}}, \bibinfo{author}{\bibfnamefont{S.}~\bibnamefont{Shova}}, \bibinfo{author}{\bibfnamefont{L.}~\bibnamefont{Chioncel}}, \bibinfo{author}{\bibfnamefont{V.}~\bibnamefont{Tsurkan}}, \bibnamefont{et~al.}, \bibinfo{journal}{Applied Physics Letters} \textbf{\bibinfo{volume}{123}} (\bibinfo{year}{2023}).

\bibitem[{\citenamefont{Kurosawa et~al.}(2024)\citenamefont{Kurosawa, Higo, Saito, Uesugi, and Nakatsuji}}]{kurosawa2024large}
\bibinfo{author}{\bibfnamefont{S.}~\bibnamefont{Kurosawa}}, \bibinfo{author}{\bibfnamefont{T.}~\bibnamefont{Higo}}, \bibinfo{author}{\bibfnamefont{S.}~\bibnamefont{Saito}}, \bibinfo{author}{\bibfnamefont{R.}~\bibnamefont{Uesugi}}, \bibnamefont{and} \bibinfo{author}{\bibfnamefont{S.}~\bibnamefont{Nakatsuji}}, \bibinfo{journal}{Physical Review Materials} \textbf{\bibinfo{volume}{8}}, \bibinfo{pages}{054206} (\bibinfo{year}{2024}).

\bibitem[{\citenamefont{Fu et~al.}(2024)\citenamefont{Fu, Yu, Zhang, Li, and Liu}}]{fu2024magnetic}
\bibinfo{author}{\bibfnamefont{X.}~\bibnamefont{Fu}}, \bibinfo{author}{\bibfnamefont{J.}~\bibnamefont{Yu}}, \bibinfo{author}{\bibfnamefont{Q.}~\bibnamefont{Zhang}}, \bibinfo{author}{\bibfnamefont{Z.}~\bibnamefont{Li}}, \bibnamefont{and} \bibinfo{author}{\bibfnamefont{Z.}~\bibnamefont{Liu}}, \bibinfo{journal}{Journal of Physics and Chemistry of Solids} p. \bibinfo{pages}{112473} (\bibinfo{year}{2024}).

\bibitem[{\citenamefont{Prodan et~al.}(2024)\citenamefont{Prodan, Chmeruk, Chioncel, Tsurkan, and K{\'e}zsm{\'a}rki}}]{prodan2024anisotropic}
\bibinfo{author}{\bibfnamefont{L.}~\bibnamefont{Prodan}}, \bibinfo{author}{\bibfnamefont{A.}~\bibnamefont{Chmeruk}}, \bibinfo{author}{\bibfnamefont{L.}~\bibnamefont{Chioncel}}, \bibinfo{author}{\bibfnamefont{V.}~\bibnamefont{Tsurkan}}, \bibnamefont{and} \bibinfo{author}{\bibfnamefont{I.}~\bibnamefont{K{\'e}zsm{\'a}rki}}, \bibinfo{journal}{Physical Review B} \textbf{\bibinfo{volume}{110}}, \bibinfo{pages}{094407} (\bibinfo{year}{2024}).

\bibitem[{\citenamefont{Li et~al.}(2022)\citenamefont{Li, Zhang, and Hong}}]{li2022anisotropic}
\bibinfo{author}{\bibfnamefont{T.}~\bibnamefont{Li}}, \bibinfo{author}{\bibfnamefont{L.}~\bibnamefont{Zhang}}, \bibnamefont{and} \bibinfo{author}{\bibfnamefont{X.}~\bibnamefont{Hong}}, \bibinfo{journal}{Journal of Vacuum Science \& Technology A} \textbf{\bibinfo{volume}{40}} (\bibinfo{year}{2022}).

\bibitem[{\citenamefont{Smit}(1955)}]{smit1955spontaneous}
\bibinfo{author}{\bibfnamefont{J.}~\bibnamefont{Smit}}, \bibinfo{journal}{Physica} \textbf{\bibinfo{volume}{21}}, \bibinfo{pages}{877} (\bibinfo{year}{1955}).

\bibitem[{\citenamefont{Smit}(1958)}]{smit1958spontaneous}
\bibinfo{author}{\bibfnamefont{J.}~\bibnamefont{Smit}}, \bibinfo{journal}{Physica} \textbf{\bibinfo{volume}{24}}, \bibinfo{pages}{39} (\bibinfo{year}{1958}).

\bibitem[{\citenamefont{Berger}(1970)}]{berger1970side}
\bibinfo{author}{\bibfnamefont{L.}~\bibnamefont{Berger}}, \bibinfo{journal}{Physical Review B} \textbf{\bibinfo{volume}{2}}, \bibinfo{pages}{4559} (\bibinfo{year}{1970}).

\bibitem[{\citenamefont{Hellman et~al.}(2017)\citenamefont{Hellman, Hoffmann, Tserkovnyak, Beach, Fullerton, Leighton, MacDonald, Ralph, Arena, D{\"u}rr et~al.}}]{hellman2017interface}
\bibinfo{author}{\bibfnamefont{F.}~\bibnamefont{Hellman}}, \bibinfo{author}{\bibfnamefont{A.}~\bibnamefont{Hoffmann}}, \bibinfo{author}{\bibfnamefont{Y.}~\bibnamefont{Tserkovnyak}}, \bibinfo{author}{\bibfnamefont{G.~S.} \bibnamefont{Beach}}, \bibinfo{author}{\bibfnamefont{E.~E.} \bibnamefont{Fullerton}}, \bibinfo{author}{\bibfnamefont{C.}~\bibnamefont{Leighton}}, \bibinfo{author}{\bibfnamefont{A.~H.} \bibnamefont{MacDonald}}, \bibinfo{author}{\bibfnamefont{D.~C.} \bibnamefont{Ralph}}, \bibinfo{author}{\bibfnamefont{D.~A.} \bibnamefont{Arena}}, \bibinfo{author}{\bibfnamefont{H.~A.} \bibnamefont{D{\"u}rr}}, \bibnamefont{et~al.}, \bibinfo{journal}{Reviews of modern physics} \textbf{\bibinfo{volume}{89}}, \bibinfo{pages}{025006} (\bibinfo{year}{2017}).

\bibitem[{\citenamefont{Drijver et~al.}(1976)\citenamefont{Drijver, Sinnema, and Van~der Woude}}]{drijver1976magnetic}
\bibinfo{author}{\bibfnamefont{J.}~\bibnamefont{Drijver}}, \bibinfo{author}{\bibfnamefont{S.}~\bibnamefont{Sinnema}}, \bibnamefont{and} \bibinfo{author}{\bibfnamefont{F.}~\bibnamefont{Van~der Woude}}, \bibinfo{journal}{Journal of Physics F: Metal Physics} \textbf{\bibinfo{volume}{6}}, \bibinfo{pages}{2165} (\bibinfo{year}{1976}).

\bibitem[{\citenamefont{Lou et~al.}(2024)\citenamefont{Lou, Zhou, Song, Fedorov, Tu, Jiang, Wang, Li, Liu, Chen et~al.}}]{lou2024orbital}
\bibinfo{author}{\bibfnamefont{R.}~\bibnamefont{Lou}}, \bibinfo{author}{\bibfnamefont{L.}~\bibnamefont{Zhou}}, \bibinfo{author}{\bibfnamefont{W.}~\bibnamefont{Song}}, \bibinfo{author}{\bibfnamefont{A.}~\bibnamefont{Fedorov}}, \bibinfo{author}{\bibfnamefont{Z.}~\bibnamefont{Tu}}, \bibinfo{author}{\bibfnamefont{B.}~\bibnamefont{Jiang}}, \bibinfo{author}{\bibfnamefont{Q.}~\bibnamefont{Wang}}, \bibinfo{author}{\bibfnamefont{M.}~\bibnamefont{Li}}, \bibinfo{author}{\bibfnamefont{Z.}~\bibnamefont{Liu}}, \bibinfo{author}{\bibfnamefont{X.}~\bibnamefont{Chen}}, \bibnamefont{et~al.}, \bibinfo{journal}{Nature Communications} \textbf{\bibinfo{volume}{15}}, \bibinfo{pages}{9823} (\bibinfo{year}{2024}).

\bibitem[{\citenamefont{Wang et~al.}(2018)\citenamefont{Wang, Xu, Lou, Liu, Li, Huang, Shen, Weng, Wang, and Lei}}]{wang2018large}
\bibinfo{author}{\bibfnamefont{Q.}~\bibnamefont{Wang}}, \bibinfo{author}{\bibfnamefont{Y.}~\bibnamefont{Xu}}, \bibinfo{author}{\bibfnamefont{R.}~\bibnamefont{Lou}}, \bibinfo{author}{\bibfnamefont{Z.}~\bibnamefont{Liu}}, \bibinfo{author}{\bibfnamefont{M.}~\bibnamefont{Li}}, \bibinfo{author}{\bibfnamefont{Y.}~\bibnamefont{Huang}}, \bibinfo{author}{\bibfnamefont{D.}~\bibnamefont{Shen}}, \bibinfo{author}{\bibfnamefont{H.}~\bibnamefont{Weng}}, \bibinfo{author}{\bibfnamefont{S.}~\bibnamefont{Wang}}, \bibnamefont{and} \bibinfo{author}{\bibfnamefont{H.}~\bibnamefont{Lei}}, \bibinfo{journal}{Nature communications} \textbf{\bibinfo{volume}{9}}, \bibinfo{pages}{1} (\bibinfo{year}{2018}).

\bibitem[{\citenamefont{Cheng et~al.}(2022)\citenamefont{Cheng, Wang, Lyalin, Bagu{\'e}s, Bishop, McComb, and Kawakami}}]{cheng2022atomic}
\bibinfo{author}{\bibfnamefont{S.}~\bibnamefont{Cheng}}, \bibinfo{author}{\bibfnamefont{B.}~\bibnamefont{Wang}}, \bibinfo{author}{\bibfnamefont{I.}~\bibnamefont{Lyalin}}, \bibinfo{author}{\bibfnamefont{N.}~\bibnamefont{Bagu{\'e}s}}, \bibinfo{author}{\bibfnamefont{A.~J.} \bibnamefont{Bishop}}, \bibinfo{author}{\bibfnamefont{D.~W.} \bibnamefont{McComb}}, \bibnamefont{and} \bibinfo{author}{\bibfnamefont{R.~K.} \bibnamefont{Kawakami}}, \bibinfo{journal}{APL Materials} \textbf{\bibinfo{volume}{10}} (\bibinfo{year}{2022}).

\bibitem[{\citenamefont{Kresse and Furthm{\"u}ller}(1996{\natexlab{a}})}]{kresse1996efficiency}
\bibinfo{author}{\bibfnamefont{G.}~\bibnamefont{Kresse}} \bibnamefont{and} \bibinfo{author}{\bibfnamefont{J.}~\bibnamefont{Furthm{\"u}ller}}, \bibinfo{journal}{Computational materials science} \textbf{\bibinfo{volume}{6}}, \bibinfo{pages}{15} (\bibinfo{year}{1996}{\natexlab{a}}).

\bibitem[{\citenamefont{Kresse and Furthm{\"u}ller}(1996{\natexlab{b}})}]{kresse1996efficient}
\bibinfo{author}{\bibfnamefont{G.}~\bibnamefont{Kresse}} \bibnamefont{and} \bibinfo{author}{\bibfnamefont{J.}~\bibnamefont{Furthm{\"u}ller}}, \bibinfo{journal}{Physical review B} \textbf{\bibinfo{volume}{54}}, \bibinfo{pages}{11169} (\bibinfo{year}{1996}{\natexlab{b}}).

\bibitem[{\citenamefont{Perdew et~al.}(1996)\citenamefont{Perdew, Burke, and Ernzerhof}}]{perdew1996generalized}
\bibinfo{author}{\bibfnamefont{J.~P.} \bibnamefont{Perdew}}, \bibinfo{author}{\bibfnamefont{K.}~\bibnamefont{Burke}}, \bibnamefont{and} \bibinfo{author}{\bibfnamefont{M.}~\bibnamefont{Ernzerhof}}, \bibinfo{journal}{Physical review letters} \textbf{\bibinfo{volume}{77}}, \bibinfo{pages}{3865} (\bibinfo{year}{1996}).

\bibitem[{\citenamefont{Mostofi et~al.}(2008)\citenamefont{Mostofi, Yates, Lee, Souza, Vanderbilt, and Marzari}}]{mostofi2008wannier90}
\bibinfo{author}{\bibfnamefont{A.~A.} \bibnamefont{Mostofi}}, \bibinfo{author}{\bibfnamefont{J.~R.} \bibnamefont{Yates}}, \bibinfo{author}{\bibfnamefont{Y.-S.} \bibnamefont{Lee}}, \bibinfo{author}{\bibfnamefont{I.}~\bibnamefont{Souza}}, \bibinfo{author}{\bibfnamefont{D.}~\bibnamefont{Vanderbilt}}, \bibnamefont{and} \bibinfo{author}{\bibfnamefont{N.}~\bibnamefont{Marzari}}, \bibinfo{journal}{Computer physics communications} \textbf{\bibinfo{volume}{178}}, \bibinfo{pages}{685} (\bibinfo{year}{2008}).

\bibitem[{\citenamefont{Tsirkin}(2021)}]{tsirkin2021high}
\bibinfo{author}{\bibfnamefont{S.~S.} \bibnamefont{Tsirkin}}, \bibinfo{journal}{npj Computational Materials} \textbf{\bibinfo{volume}{7}}, \bibinfo{pages}{33} (\bibinfo{year}{2021}).

\bibitem[{\citenamefont{Nagaosa et~al.}(2010{\natexlab{b}})\citenamefont{Nagaosa, Sinova, Onoda, MacDonald, and Ong}}]{nagaosa2010ahe}
\bibinfo{author}{\bibfnamefont{N.}~\bibnamefont{Nagaosa}}, \bibinfo{author}{\bibfnamefont{J.}~\bibnamefont{Sinova}}, \bibinfo{author}{\bibfnamefont{S.}~\bibnamefont{Onoda}}, \bibinfo{author}{\bibfnamefont{A.~H.} \bibnamefont{MacDonald}}, \bibnamefont{and} \bibinfo{author}{\bibfnamefont{N.~P.} \bibnamefont{Ong}}, \bibinfo{journal}{Rev. Mod. Phys.} \textbf{\bibinfo{volume}{82}}, \bibinfo{pages}{1539} (\bibinfo{year}{2010}{\natexlab{b}}).

\bibitem[{\citenamefont{Xiao et~al.}(2010)\citenamefont{Xiao, Chang, and Niu}}]{xiao2010berry}
\bibinfo{author}{\bibfnamefont{D.}~\bibnamefont{Xiao}}, \bibinfo{author}{\bibfnamefont{M.-C.} \bibnamefont{Chang}}, \bibnamefont{and} \bibinfo{author}{\bibfnamefont{Q.}~\bibnamefont{Niu}}, \bibinfo{journal}{Rev. Mod. Phys.} \textbf{\bibinfo{volume}{82}}, \bibinfo{pages}{1959} (\bibinfo{year}{2010}).

\end{thebibliography}

\section{Figures}
\begin{figure}[H]
    \centering
    \includegraphics[width=1\linewidth]{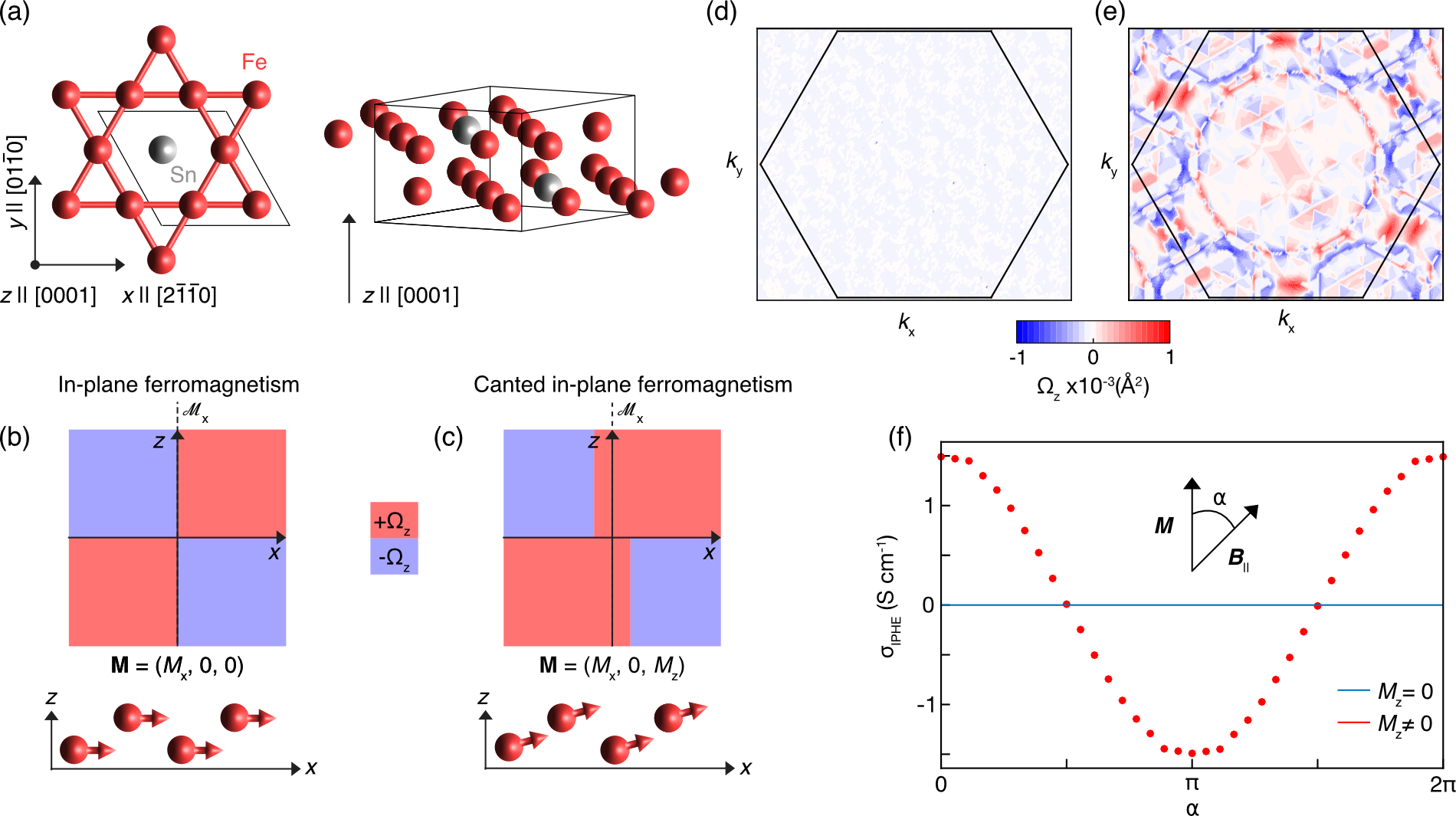}
    \caption{{\bf Prediction of the anomalous Hall in-plane Hall effect in the Weyl ferromagnet Fe$_3$Sn.} (a) Left panel: Shown is the top view of the lattice structure of Fe$_3$Sn within the two-dimensional kagome plane ($xy$-plane). The hexagonal unit cell (black solid line) and definition of Cartesian coordinates with respect to the crystallographic axes are indicated. Right panel: Shown is an isometric view of the bilayer stacking of the kagome layer along the $z$-direction. (b) and (c), shown are the schematic distributions of the total Berry curvature within the $xz$-plane at $k_{\rm y}=0$ for an ideal in-plane ferromagnet with magnetization vector ${\bf M}=(M_{\rm x},\,0,\,0)$ and an in-plane ferromagnet with out-of-plane canting ${\bf M}=(M_{\rm x},\,0,\,M_{\rm z})$, respectively. The mirror plane $\mathcal{M}_{\rm x}$ is indicated. In case without canting, $\mathcal{M}_{\rm z}t\mathcal{T}$ is preserved and the total Berry curvature integrates to zero. A finite spin canting $M_{\rm z}$ breaks $\mathcal{M}_{\rm z}t\mathcal{T}$ resulting in a finite total Berry curvature. (d) and (e), shown are the energy-integrated calculated Berry curvature obtained from a Wannierized {\em ab-initio} band structure of Fe$_3$Sn within the first hexagonal Brioullin zone (black solid line) at $k_{\rm z}=0$ with and without out-of-plane spin canting, respectively. (f) Shown is the calculated anomalous in-plane Hall conductivity $\sigma_{\rm IPHE}(\alpha)$ as a function of the angle $\alpha$ between ${\bf M}$ and the direction of an in-plane magnetic field $\bf B_{\parallel}$ as schematically indicated. Without canting, $\hat{{\bf M}}=(1,\,0,\,0)$, the magnetic symmetry $\mathcal{M}_{\rm z}t\mathcal{T}$ is preserved, resulting in $\sigma_{\rm IPHE}(\alpha)=0$ (blue solid line). Breaking of this symmetry by a finite out-of-plane canting $\hat{{\bf M}}=(1,\,0,\,0.1)$ permits a finite $\sigma_{\rm IPHE}(\alpha)$ (solid red dots). Details of the model calculations are presented in the Methods section.}
    \label{fig:fig1}
\end{figure}
\clearpage
\begin{figure}[H]
    \centering
    \includegraphics[width=1\linewidth]{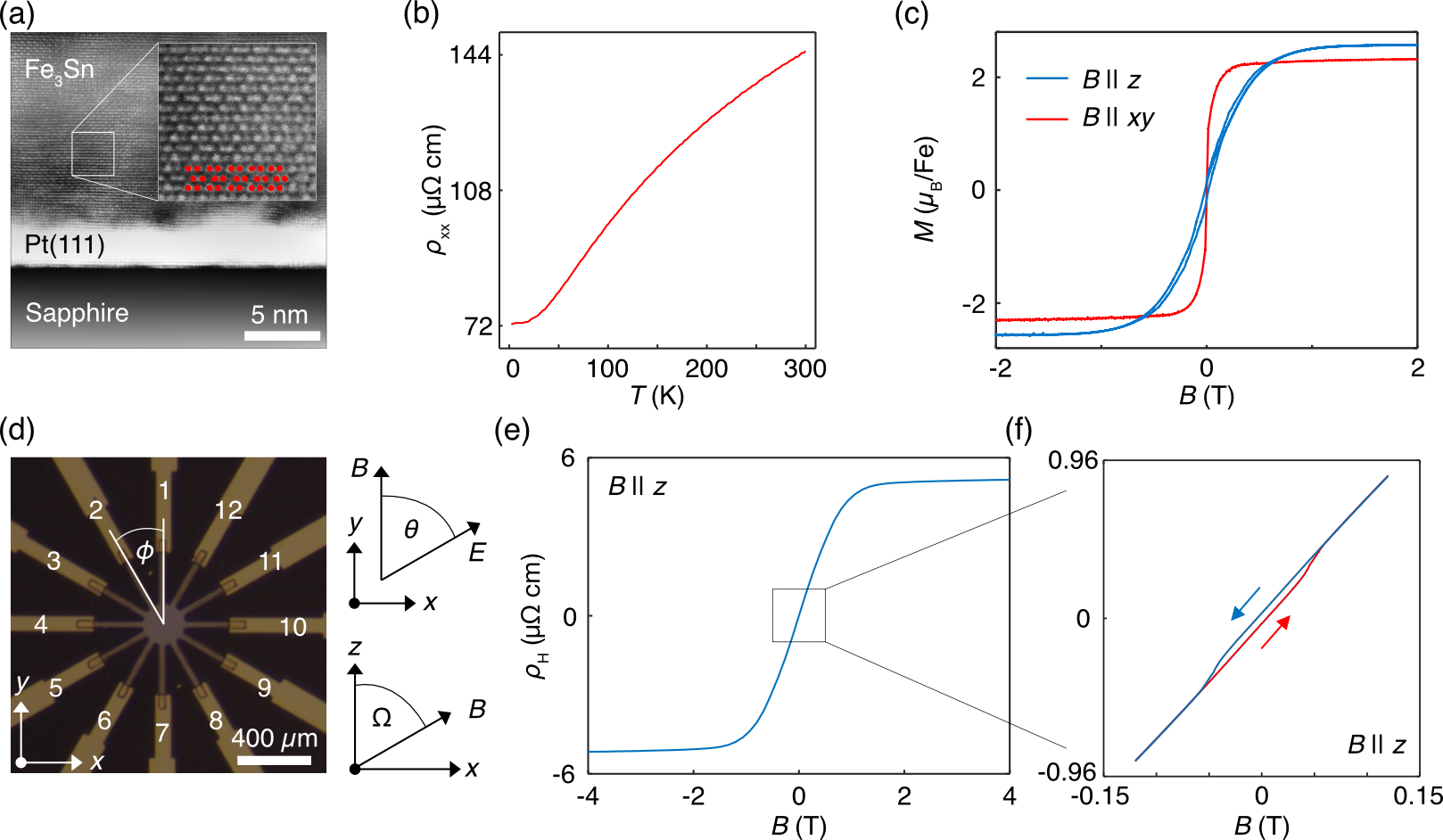}
    \caption{{\bf Detection of spin-canted in-plane ferromagnetism in Fe$_3$Sn thin films.} (a) Shown is a transmission electron microscopy cross-sectional image of the Fe$_3$Sn(001)/Pt(111)/sapphire(0001) thin film within the $[11\bar{2}0]-[0001]$-plane. The different layers are annotated. The inset shows a magnification of the atomic layer structure of the Fe$_3$Sn layers and highlights their bilayer stacking along the $z$-direction. Fe atoms are schematically indicated by red spheres. (b) Shown is the temperature ($T$) dependence of the longitudinal resistivity $\rho_{\rm xx}$ (solid red line) of Fe$_3$Sn\,(60\,nm)/Pt\,(5\,nm)/sapphire. (c) Shown is the magnetization $M$ of Fe$_3$Sn\,(30\,nm)/Pt\,(5\,nm)/sapphire for magnetic fields $B$ applied within the $xy$-plane (red solid line) and along the $z$-direction (blue solid line) at $T=300\,$K. (d) Optical micrograph of the circular device structure fabricated from the Fe$_3$Sn/Pt(111)/sapphire(0001) thin films. The numerical contact labels are indicated. The azimuthal (within $xy$-plane) angle $\phi$ is measured between the direction of the electric field $\bf E$ and the $y$-direction of the sample in counterclockwise direction. The azimuthal angle $\theta$ is defined as the angle between $\bf E$ and the in-plane magnetic field $\bf B_{\parallel}$ in counterclockwise direction. Here, $\bf E$ is always parallel to the direction of the applied electric bias current $I$. The polar angle $\Omega$ is defined as the angle between the out-of-plane component of the external magnetic field $\bf B$ and the $z$-direction. (e) Shown is the Hall resistivity $\rho_{\rm H}$ at $\phi=0$ measured as a function of $B$ applied along the $z-$direction ($\Omega=0$) at $T=300\,$K. (f) Shown is a magnification of $\rho_{\rm H}$ in panel e at small magnetic field amplitudes for upward and downward sweeps of $B$ recorded at $\phi=0$ and $T=300\,$K.}
    \label{fig:fig2}
\end{figure}

\begin{figure}[H]
    \centering
    \includegraphics[width=1\linewidth]{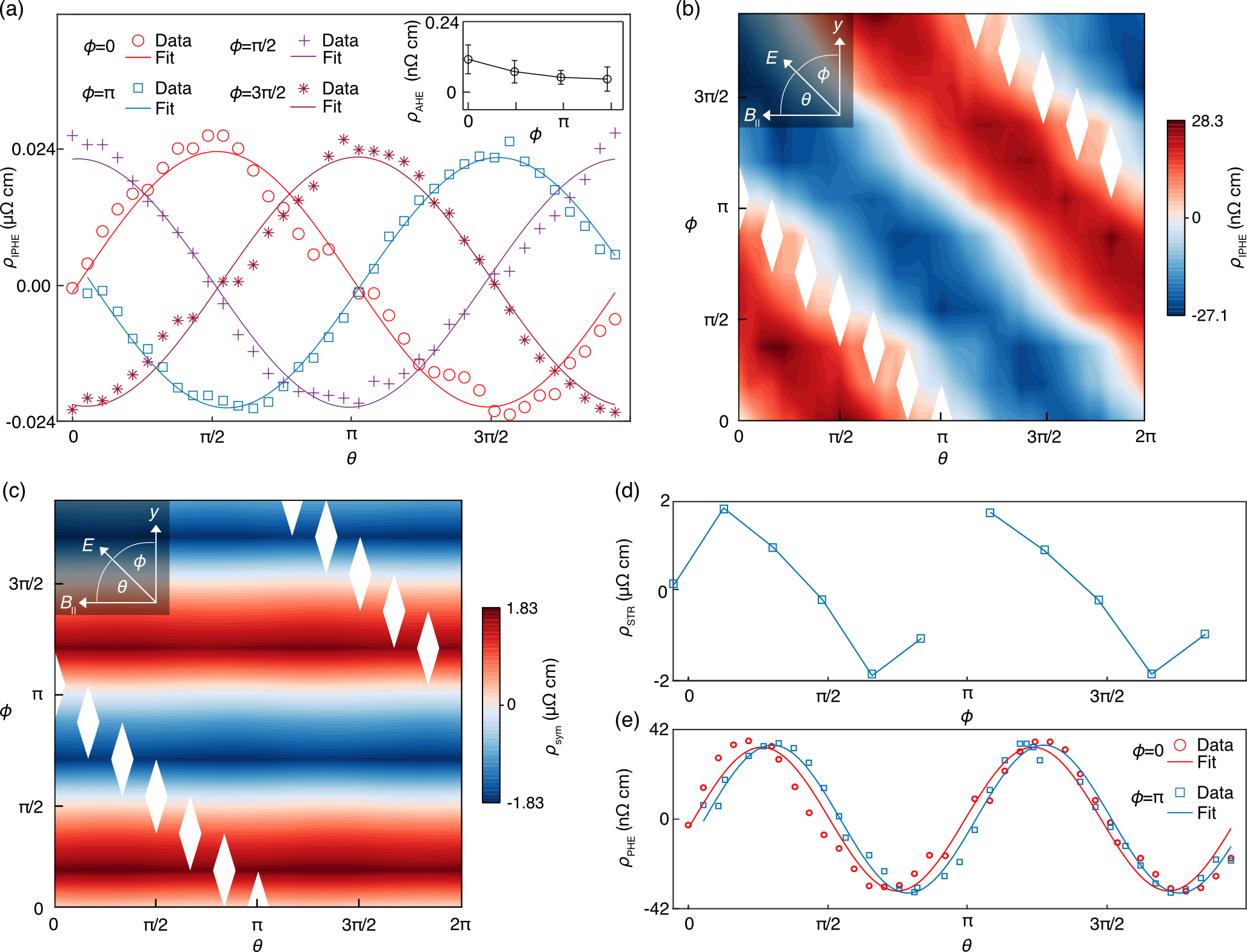}
    \caption{{\bf Experimental observation of the anomalous in-plane Hall effect in Fe$_3$Sn at room temperature.} (a) Shown is the in-plane Hall resistivity $\rho_{\rm IPHE}(\theta)$ as a function of the angle $\theta$ recorded on device D1 at different, indicated values of the angle $\phi$ at a temperature $T=300\,$K and an in-plane magnetic field $B_{\parallel}=20\,$mT ($\Omega=\pi/2$). Shown are the data (symbols) and fit to data (solid lines). The inset shows the anomalous Hall resistivity $\rho_{\rm AHE}$ obtained from fits to the data at different values of $\phi$ (see main text). (b) Shown is the full angle dependence of $\rho_{\rm IPHE}(\phi,\,\theta)$ within the $\phi-\theta$-plane recorded on device D1 at a temperature $T=300\,$K and an in-plane magnetic field $B_{\parallel}=20\,$mT ($\Omega=\pi/2$). (c) Shown is the full angle dependence of the symmetric contribution $\rho_{\rm sym}(\phi,\,\theta)$ to the transverse resistivity within the $\phi-\theta$-plane recorded on device D1 at a temperature $T=300\,$K and an in-plane magnetic field $B_{\parallel}=20\,$mT ($\Omega=\pi/2$). (d) Shown is the symmetric transverse resistivity $\rho_{\rm STR}$ recorded as a function of the angle $\phi$ on device D1 at $\theta=0$, $T=300\,$K, and $B_{\parallel}=20\,$mT ($\Omega=\pi/2$). (e) Shown is the resistivity $\rho_{\rm PHE}$ of the planar Hall effect (PHE) recorded as a function of $\theta$ on device D1 at $\phi=0$ and $\phi=\pi$ at $T=300\,$K, and $B_{\parallel}=20\,$mT ($\Omega=\pi/2$). Shown are the data (symbols) and fit to data (solid lines) with fitted amplitudes $\rho_{\rm PHE}(\phi=0)=(33.6\pm1.7)\,\text{n}\Omega\,\text{cm}$ and $\rho_{\rm PHE}(\phi=\pi)=(34.6\pm0.9)\,\text{n}\Omega\,\text{cm}$. {\em Note: The two-dimensional resistivity maps in panels b and c result from interpolating $\theta$-dependent resistivity curves recorded at twelve different values of the angle $\phi$. The white diamonds indicate missing data points (also seen in panel d and e) that could not be reached owing to the travel limit of the rotator probe (see Methods section).}}
    \label{fig:fig3}
\end{figure}

\begin{figure}[H]
    \centering
    \includegraphics[width=1\linewidth]{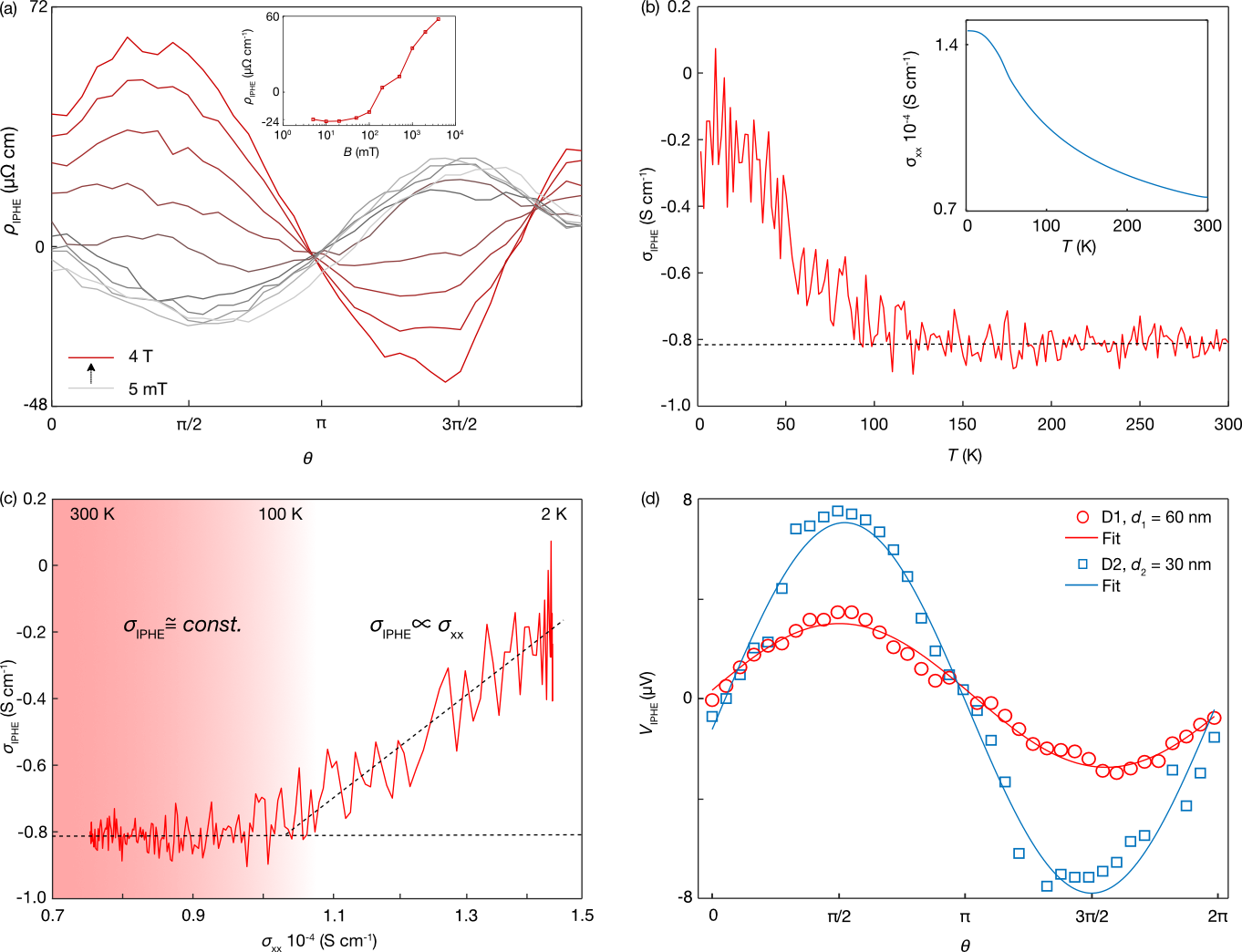}
    \caption{{\bf Origin of the anomalous in-plane Hall effect in Fe$_3$Sn} (a) Shown is the in-plane Hall resistivity $\rho_{\rm IPHE}$ of device D1 plotted as a function of the angle $\theta$ recorded at $\phi=0$ and a temperature $T=300\,$K for different amplitudes of the magnetic field $\bf B_{\parallel}$ (see legend). The inset shows the amplitude of $\rho_{\rm IPHE}$ at $\theta=\pi/2$ plotted as a function of $\bf B$ as extracted from the data in the main panel. (b) Shown is the temperature dependence of the anomalous in-plane Hall conductivity $\sigma_{\rm IPHE}$ of device D1 recorded at $B_{\parallel}=50\,$mT, $\phi=\pi$, and $\theta=\pi/2$. The inset displays the temperature dependence of the corresponding longitudinal conductivity $\sigma_{\rm xx}(T)$. (c) Shown is the dependence of $\sigma_{\rm IPHE}$ on $\sigma_{\rm xx}$ of the data shown in panel b. The different temperature ranges and conductivity regimes are indicated. (d) Shown is the $\theta$ dependence of the magnetic field-asymmetric part $V_{\rm IPHE}(\theta)$ of the measured transverse voltage $V_{\rm xy}(\theta)$ recorded on devices D1 ($d_{\rm1}=60\,$nm) and D2 ($d_{\rm2}=30\,$nm) at $T=300\,$K, $B_{\parallel}=20\,$mT, $\Omega=\pi/2$, and $\phi=0$. Data are shown as symbols, and fits to the data are shown as solid lines. The fitted voltage amplitudes are $V_{\rm IPHE}(60\,\text{nm})=(2.9\pm0.1)\,\mu$V and $V_{\rm IPHE}(30\,\text{nm})=(7.4\pm0.2)\,\mu$V.}
    \label{fig:fig4}
\end{figure}

\begin{figure}[H]
    \centering
    \includegraphics[width=1\linewidth]{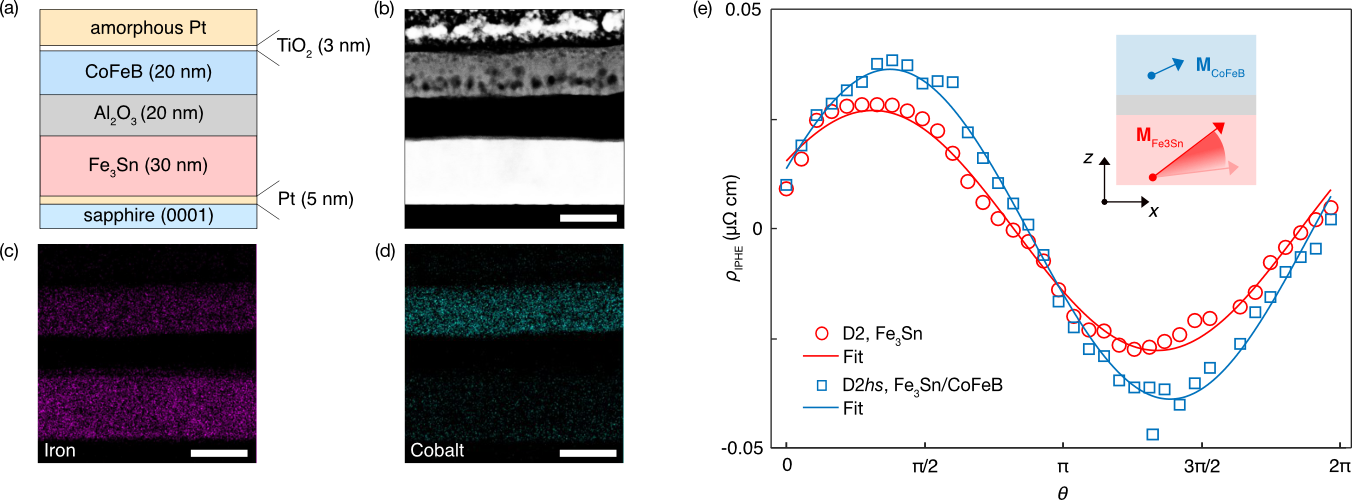}
    \caption{{\bf Tuning the anomalous IPHE in a topological heterostructure.} (a) Shown is a schematic overview of the layer structure and composition of the topological heterostructure device D2{\em hs}. (b) Shown is a corresponding cross-sectional scanning transmission electron microscopy (STEM) image of an 100\,nm$\times$100\,nm area D2{\em hs}. The scale bar corresponds to a length of 25\,nm. The individual layers of the multilayer structure can be distinguished by their different black-white contrast. (c) and (d) display the elemental distribution of iron and cobalt, respectively in the same field of view as panel b recorded using energy-dispersive X-ray spectroscopy (EDS) with the STEM. A more detailed characterization of the topological heterostructure is presented in Sec.~VII of the Suppl.~Materials. (e) Shown is a the $\theta$ dependence of the anomalous in-plane Hall resistivity $\rho_{\rm IPHE}$ (symbols) and fit to the data (solid lines) recorded on device D2 and the heterostructure device D2{\em hs} at $T=300\,$K, $B_{\parallel}=20\,$mT, $\Omega=\pi/2$, and $\phi=0$. The fitted amplitudes are $\rho^0_{\rm IPHE}(\text{D2})=(26.4\pm0.6)\,\text{n}\Omega\,\text{cm}$ and $\rho^0_{\rm IPHE}(\text{D2{\em hs}})=(36.3\pm0.6)\,\text{n}\Omega\,\text{cm}$. The heterostructure and the effect of the magnetization $\bf M$ of CoFeB on the magnetization of Fe$_3$Sn via its magnetic stray field are schematically indicated.}
    \label{fig:fig5}
\end{figure}

\newpage
\section{Methods}

\subsection{Molecular beam epitaxy of Fe$_{3}$Sn films}
Epitaxial Fe$_{3}$Sn films were grown on top of a 5\,nm thick Pt(111) buffer layer on $[0001]$-oriented 5\,mm x 5\,mm x 0.5\,mm sized sapphire substrates (from CrysTec GmbH) using a home-build molecular beam epitaxy (MBE) system (effusion cells from MBE Komponenten GmbH). The as-received sapphire substrates were chemically cleaned using sonication in acetone and isopropyl alcohol (IPA) for 5\,min each. Atomically flat surfaces with single-step height terraces were obtained by thermal annealing in air at 1300\,K for 1\,h using a tube furnace. Prior to the thin-film deposition, the substrates were degassed inside the MBE chamber at 900\,K at a base pressure $p\leq5\times10^{-10}$\,mbar. Adopting the recipe of Cheng {\em et al.}~\cite{cheng2022atomic}, high purity Pt(99.999\,$\%$) was evaporated using an electron beam evaporator (EFM4 from Focus GmbH) at an approximate rate of 0.2\,{\AA}/min to obtain a 5\,nm thick Pt layer oriented along the $[111]$-direction. The initial 6\,{\AA} of Pt were deposited at a substrate temperature of 740\,K. The remaining 4.4\,{\AA} were deposited while the substrate temperature was reduced from 440\,K to 380\,K, followed by annealing at 600\,K for 10\,min. Onward, high purity Fe(99.999\%) and Sn(99.999\%) were co-evaporated using effusion cells at an approximate growth rate of 1.2\,{\AA}/min to obtain Fe$_3$Sn films. The structural and chemical characterization of the fabricated thin films is described in Sec.~II of the Suppl.~Materials.

\subsection{Device fabrication and deposition of topological heterostructure}
The as-grown Fe$_3$Sn thin films were shaped into 12-terminal circular Hall bar devices using Ar$^+$ ion milling and optical UV lithography. The electrical contacts to the Hall bars were fabricated by evaporating 5\,nm/100\,nm titanium/gold electrodes. The heterostructure device D2{\em hs} was prepared directly on top of the Fe$_3$Sn circular hall bars of device D2 using a UV lithography mask. An e-beam evaporator (Innovative Vaccum System) was used to deposit 20\,nm of amorphous Al$_{2}$O$_{3}$. In the next step, a 20\,nm thick amorphous CoFeB layer was deposited using DC magnetron-sputtering (AJA International Ltd.) at a base pressure of 1x10$^{-8}$\,torr with a 3\,mtorr partial pressure of argon and a d.c. power of 60\,W. Following CoFeB deposition, a 3\,nm thick titanium (Ti) capping layer was sputter-deposited to protect the heterostructure from oxidation, where the Ti layer naturally oxidizes into TiO$_2$ via the exposure to an ambient atmosphere. The material characterization of the CoFeB films is described in Sec.~VII of the Suppl.~Materials.

\subsection{Electric transport measurements}
All electric transport measurements were carried out in a Quantum Design PPMS 6000 system using a four-probe contact geometry. The circular Hall bar devices were wire bound to the chip holder using 25\,$\mu$m thick aluminum wire. An a.c. bias current of $I=I_{\rm 0}\sin(\omega t)$ with a peak amplitude of $I_{\rm 0}=1.1\,$mA at a frequency of $f=\omega/2\pi=19.375\,$Hz was applied by using an a.c. current source (6221A, Keithley). Two lock-in amplifiers (SRS 830, Stanford Research Systems), which are phase matched to the bias current output, were used to record the longitudinal $V_{\rm xx}$ and transverse $V_{\rm xy}$ voltages. The angle $\theta$ between the electric field $\bf E$ and the in-plane magnetic field $\bf B_{\parallel}$, as defined in Fig.~\ref{fig:fig2}(d), was controlled using a commercial rotor probe insert from Quantum Design. Note that the angle $\theta$ can be tuned from $\theta=0$° to only $\theta=350$° owing to the travel limitation of the rotator probe. This causes the missing data points at $\theta=360\text{°}=2\pi$ as seen in Fig.~\ref{fig:fig3}. The longitudinal and transverse resistivities were obtained by using the relations $\rho_{\rm xx}=(d_{\rm 1,2}L_{\rm y}/L_{\rm x})(V_{\rm xx}/I_{\rm x})$ and $\rho_{\rm xy}=(d_{\rm 1,2}L_{\rm y}/L_{\rm x})(V_{\rm xy}/I_{\rm x})$ with the device dimension, $L_{\rm x}=L_{\rm y}=200\,\mu$m and thicknesses $d_{\rm 1}=60\,$nm and $d_{\rm 2}=30\,$nm respectively.

\subsection{Magnetization measurements}
Measurements of sample magnetization were performed using a vibrating sample magnetometer (MPMS3 from Quantum Design, resolution $<10^{-8}\,$emu).  The samples were attached to standard quartz or brass holders using GE varnish. All measurements were performed at a temperature of 300\,K. The magnetic moment per Fe atom was calculated considering the film thickness, as determined from X-ray reflection measurements, and the effective sample area $A'$. $A'$ is smaller by 4\,mm$^2$ than the actual surface area $A=5\times5\,\text{mm}^2$ of the sapphire substrate owing to four corner clamps used to mount the substrate to the sample holder of the molecular beam epitaxy system.

\subsection{Scanning transmission electron microscopy measurements}
Lamellas suitable for carrying out scanning transmission electron microscopy (STEM) measurements were prepared using a focused ion beam system (FEI Helios G4 UX). To protect the heterostructure from potential damage caused by the Ga$^+$ion beam, an amorphous platinum (Pt) layer was deposited on the surface of the sample. Following the ion milling process, the as-prepared lamella was transferred to a copper grid.

Cross-sectional STEM measurements were performed using a commercial JEOL JEM-ARM200F system. Using an acceleration voltage of 60\,kV and a probe corrector, an image resolution below 0.1\,nm was achieved. Energy-dispersive X-ray spectroscopy measurements were performed to map out the elemental distribution in an area measuring $100\,\mu\text{m}\times100\,\mu\text{m}$.

\subsection{Density functional theory calculations}

Density functional theory (DFT) calculations of the electronic structure of Fe$_3$Sn were performed using the density functional theory framework as implemented in the Vienna {\em ab initio} simulation package~\cite{kresse1996efficiency, kresse1996efficient}. The projector-augmented wave potential was adopted with the plane-wave energy cutoff set to 600\,eV (convergence criteria $10^{-6}\,$eV). The exchange-correlation functional of the Perdew–Burke–Ernzerhof type was used~\cite{perdew1996generalized} with a 11$\times$11$\times$13 gamma-centered Monkhorst–Pack mesh. We constructed the Hamiltonian of a Wannier tight-binding model using the {\em WANNIER90} interface~\cite{mostofi2008wannier90}, including the Fe $d$-and Sn $p$-orbitals. We then calculated the anomalous Hall conductivity from this Wannier model using a 131$\times$131$\times$141 k-mesh and the {\em WannierBerri} code~\cite{tsirkin2021high}.

\subsection{Anomalous in-plane Hall effect calculated using the Wannier tight-binding model}

To analyze the characteristics of in-plane Hall effect (PHE) under an in-plane magnetic field $\bf B_{\parallel}$, we computed the anomalous in-plane Hall conductivity $\sigma_{\rm IPHE}$ using the Wannier tight-binding model projected by DFT calculations. The effective Hamiltonian is given by: 
\begin{equation*}
H=H_0+g\bf{B_{\parallel}} \cdot \boldsymbol{\sigma},
\end{equation*}
where $H_0$ refers to the Wannier Hamiltonian of Fe$_3$Sn with or without canting, the second term denotes the Zeeman coupling with ${\bf B_{\parallel}}= \rm{B} \left(\cos{\alpha},\,\sin{\alpha},\,0\right)$, with $\alpha$ as the angle between the magnetization of Fe$_3$Sn and ${\bf B_{\parallel}}$, $g$ as the effective coupling factor, and $\boldsymbol{\sigma}$ as the Pauli matrices for spin. We calculate $\sigma_{\rm xy}\equiv\sigma_{\rm IPHE}$ using the effective model through the Kubo formula~\cite{nagaosa2010ahe,xiao2010berry}:
\begin{equation}
    \sigma_{\rm xy} = \frac{e}{\hbar}\sum_{\rm n} \int_{\rm BZ} \frac{\rm{d}^3 \bf{k}}{(2\pi)^3} f_{\rm n}(\bf{k})\rm{\Omega_{n,\rm{xy}}(\bf{k})},
\end{equation}
\begin{equation}
    {\Omega_{\rm{n},\rm{xy}}(\bf{k})} = 2i\hbar^2 \sum_{\rm m\neq n} \frac{\left\langle\psi_{n\mathbf{k}}\right|\hat{v}_{\rm x}\left|\psi_{\rm{m}\bf{k}}\right\rangle \left\langle\psi_{m\mathbf{k}}\right|\hat{v}_{\rm y}\left|\psi_{\rm{n}\bf{k}}\right\rangle}{(E_{\rm{n} \bf{k}} - E_{\rm{m} \bf{k}})^2},
\end{equation}
where $f_{\rm n}\left(\bf{k}\right)$ denotes the Fermi-Dirac distribution, $\psi_{\rm{n} \bf{k}}$ and $E_{\rm{n}\bf{k}}$ refer to the Bloch wave functions and eigenvalues of band $n$, respectively, and $\hat{v}$ is the velocity operator. The calculated $\sigma_{\rm IPHE}$ using $g B_{\parallel}=0.001$\,eV is shown in Fig.~\ref{fig:fig1}(f). As expected from our symmetry analysis, $\sigma_{\rm IPHE}$ vanishes without canting, while it exhibits a finite amplitude with $2\pi$ periodicity with respect to the angle $\alpha$ when canting is included. Specifically, in the absence of canting, the magnetic space group symmetry $\mathcal{M}_zt\mathcal{T}$ is preserved under an in-plane magnetic field, resulting in opposite values of $\Omega_{\rm{n},\rm{xy}}(\bf{k})$ at symmetry related reciprocal points: $\Omega_{\rm{n},\rm{xy}}\left(\bf{k}\right)=-\Omega_{\rm{n},\rm{xy}} \left(\mathcal{M}_zt\mathcal{T} \cdot \bf{k}\right)$. Thus, $\sigma_{\rm IPHE}$ vanishes without canting, as it involves integrating $\Omega_{\rm{xy}}$ over the reciprocal space for all symmetry-determined relations of the Berry curvature (see Sec.~I of Suppl.~Materials). Meanwhile, when canting is present, $\mathcal{M}_zt\mathcal{T}$ is broken, allowing for a non-zero $\sigma_{\rm IPHE}$ based on symmetry considerations.

\section{Acknowledgments}

The authors appreciate valuable discussions with Xi Dai. This work was primarily supported by the Hong Kong Research Grant Council (Grant Nos.\,26304221, 16302422 awarded to BJ and C6033-22G awarded to BJ, JL, and RL), the Croucher Foundation (Grant No.\,CIA22SC02 awarded to BJ), and the National Key R$\&$D Program of China (Grant No.\,2021YFA1401500 awarded to JL). JL further acknowledges support from the the Hong Kong Research Grants Council (Grant Nos.\,16306722, 16304523, and C7037-22G). RL acknowledges support from the Hong Kong Research Grant Council (Grant No.\,16301022). Q.S. acknowledges support by the National Key R$\&$D Program of China (Grant No.\,2021YFA1401500). C.C. acknowledges support from the Tin Ka Ping Foundation. 

\section{Author Contributions}

BJ, JL, and SS conceived the project. SS fabricated and characterized the thin films with help from YQ and CC. SS fabricated the circular Hall bar devices and topological heterostructure and conducted the electric transport measurements with the help of YQ. XC conducted the symmetry analysis and the model calculations. SS and XC analyzed the data. TM conducted the magnetization measurements. XW deposited the CoFeB film. BJ, JL, RL and QS supervised the study. BJ wrote the manuscript with the input of all authors.

\section{Competing Interest Declaration} The authors declare that they have no competing financial interest.

\section{Data Availability Statement} Replication data for this study can be accessed on Zenodo via the link XXX.

\end{document}